\documentclass[]{jfm}
\usepackage{newtxtext}
\usepackage{newtxmath}
\usepackage{natbib}
\usepackage{hyperref}

\hypersetup{
    colorlinks = true,
    urlcolor   = blue,
    citecolor  = blue,
}

\shorttitle{Surfzone energy balance}

\shortauthor{D. P. Rijnsdorp, P. B. Smit, R. T. Guza}

\title{A nonlinear, nondispersive energy balance for surfzone waves: infragravity wave dynamics on a sloping beach}

\author{Dirk P. Rijnsdorp\aff{1},
  Pieter B. Smit\aff{2}
 \and R. T.  Guza\aff{3}}

\affiliation{\aff{1}Environmental Fluid Mechanics section, department of Hydraulic Engineering, Faculty of Civil Engineering and Geosciences, Delft University of Technology, Delft, the Netherlands
\aff{2}Sofar Ocean, San Francisco, CA, USA
\aff{3}Scripps Institution of Oceanography UCSD, La Jolla, CA, USA}

\begin{document}

\maketitle

\begin{abstract}
A fully nonlinear non-dispersive energy balance for surfzone waves is derived based on the nonlinear shallow water equations (NLSWE) to study the nearshore dynamics of infragravity (IG) waves. Based on simulations of waves on a relatively moderate and mild beach slope with a non-hydrostatic wave-flow model (SWASH), the new theory shows that spatial gradients in IG energy flux are nearly completely balanced by the combined effect of bottom stresses and predominantly nonlinear triad interactions. The new balance confirms many features of existing weakly nonlinear theories, and yields an improved description in the inner surfzone where waves become highly nonlinear. A gain of IG energy flux throughout the shoaling and outer surfzones is driven by triad interactions between IG waves and pairs of sea-swell (SS) waves. The IG energy flux decreased in the inner surfzone, primarily through an energy cascade to the swell-band and super-harmonic frequencies where wave energy is ultimately dissipated. Dissipation by bottom friction was weak on both slopes. IG wave breaking, characterized by triads between three IG or two IG waves and one SS wave, was significant only deep inside the surfzone of the mild slope. Even though IG waves broke on the mild slope, nonlinear interactions between IG waves and pairs of SS waves were responsible for at least half of the net IG flux loss.
\end{abstract}


\section{Introduction}

Linear ocean surface waves shoaling from deep to shallow water refract towards shore and steepen \citep[e.g., wavelength decreases and  amplitude increases,][]{Lamb1932}. Nonlinear processes become increasingly important in shallow water, where waves pitch forward and break \citep{Stoker1957WaterWaves}. During shoaling, nonlinear interactions cause a significant transfer of energy from the spectral peak frequency towards super and sub-harmonic frequencies \citep{Phillips1960OnInteractions,LHS1962,Hasselmann1962,Hasselmann1976AModel}. These triad interactions are off-resonant in deep and intermediate depths $kd\ge 1$ (where $k$ is the wave number and $d$ the local water depth) resulting in relatively small energy transfers. In shallow water nonlinear triads approach resonance and energy transfer towards super harmonics causes sea-swell (SS) waves to become skewed and asymmetric \citep[e.g.,][]{Elgar1985,Elgar1997SpectralBeach}, until they eventually break in the surfzone. Energy transfers to sub-harmonics excite so-called infragravity (IG) waves . The IG wave height is generally $\mathcal{O}\left(\textrm{cm}\right)$ in deep water \citep[e.g.,][]{Webb1991,Aucan2013} but can reach $\mathcal{O}\left(\textrm{m}\right)$ in shallow water during storms \citep[e.g.,][]{Sheremet2014,Matsuba2020Wave-breakingTyphoon}.

The nearshore dynamics of IG waves have been widely studied during the past decades through theoretical, laboratory, field and numerical efforts \citep[see][for a recent review]{Bertin2018InfragravityImpacts}. Various theories have been developed to explain the substantial growth of IG waves in the nearshore \citep[e.g.,][and many others]{Symonds1982,Schaffer1993a,Janssen2003,Nielsen2010,Contardo2021FreeBreaking,Liao2021AnConditions}. Such theoretical background combined with numerical modelling \citep[e.g.,][]{Reniers2002,VanDongeren2003,Lara2011}, laboratory experiments \citep[e.g.,][]{Boers1996,Baldock2000,Baldock2002}, and field campaigns \citep[e.g.,][]{Herbers1994,Herbers1995a,Herbers1995b,Okihiro1992} has significantly advanced our understanding of the IG wave patterns, growth rates, dissipation, and phase relationship with SS wave groups \citep[e.g.,][]{Battjes2004,VanDongeren2007,Baldock2012,Bakker2015}. Several studies used an IG wave energy balance to understand and quantify the growth and decay of IG waves in the nearshore \citep[e.g.,][]{Henderson2002,Henderson2006,Ruju2012,DeBakker2016}. Such theories have generally been able to explain the IG dynamics in the shoaling region and outer surfzone, but failed to provide a complete description of the dynamics in the highly nonlinear inner surfzone due to inherent assumptions of weak wave nonlinearity. In this paper, we derive a new fully nonlinear IG energy balance that aims to describe the IG wave dynamics throughout the surfzone up to the mean waterline.

Nonlinear nearshore wave dynamics have often been studied using a wave energy balance that assumes cross-shore propagation of normally incident waves \citep[e.g.,][]{Phillips1970TheOcean},
\begin{equation}
  \frac{\partial E_f}{\partial t} + \frac{\partial F_f}{\partial x} = S_f^\mathrm{NL} + D_f,
\end{equation}\label{eq:EB1}
where $t$ is time, $x$ the cross-shore coordinate, $f$ frequency, $E_f=E(x,f,t)$ the energy density spectrum, $F_f=F(x,f,t)$ the energy flux, $S_f^\mathrm{NL}=S(x,f,t)$ represents the nonlinear interactions that conservatively distribute energy over frequencies, and $D_f=D(x,f,t)$ accounts for the dissipation of wave energy (e.g., due to wave breaking and bottom friction). Terms in the wave energy balance have often been evaluated using a perturbation approach where wave nonlinearity ($\delta=\frac{a}{d}$, where $a$ is the wave amplitude) and dispersive effects ($\mu=\frac{d}{L}$, where $L$ is the wave length) are accounted for up to some order \citep[e.g.,][]{Freilich1984NonlinearWaves,Agnon1997StochasticSpectra,Herbers1997}. For example, \citet{Herbers2000} studied the spectral energy balance of breaking SS waves by evaluating the flux from linear theory $F_f=\rho g c_g E_f$ (with $c_g$ the wave group velocity, $g$ the gravitational acceleration, and $\rho$ the water density) and with a nonlinear interaction term $S^\mathrm{NL}$ from classical Boussinesq scaling, $\delta=\mathcal{O}\left(\mu^2\right)$. A perturbation approach necessarily truncates solutions at finite order and are formally invalid (no series convergence) when $\delta=O(1)$. Furthermore, energy balances (or stochastic wave models) derived using the perturbation approach depend on an infinite hierarchy of higher order moments (bi-spectrum, tri-spectrum, etc.) without natural closure \citep[e.g.,][]{Smit2015}.

Nearshore IG wave dynamics have been examined with various approximate wave energy balances. Many studies assumed small nonlinear contributions, and estimated the IG flux from linear theory \citep[e.g.,][]{Thomson2006,VanDongeren2007,TorresFreyermuth2010,Bakker2015,Liao2021AnConditions}. A subset of nonlinear contributions to the energy flux can be included by using a fully nonlinear low-frequency energy balance \citep[e.g.,][]{Phillips1970TheOcean,Schaffer1993a}, which assumes that IG frequencies are much lower than SS frequencies. Nonlinear contributions from correlations between three IG wave components are included, but (typically stronger) correlations between combinations of IG and SS components are not accounted for. \citet{Henderson2006} derived a weakly nonlinear energy balance that includes correlations between one IG and two SS components, the interaction that drives the off-resonant "bound" IG wave observed well seaward of the surfzone \citep{LHS1962}. These triads become near-resonant during SS shoaling and breaking, and have been shown to explain much of the observed cross-shore variation of IG energy flux \citep[e.g.,][]{Henderson2006,Ruju2012,Guedes2013,Rijnsdorp2015a,Mendes2018InfragravityStudy}. \citet{Bakker2015,DeBakker2016} used $S^\mathrm{NL}$ from Boussinesq scaling to show that nonlinear interactions between two IG and one SS component can become significant and must be included to explain the loss of IG flux near the shoreline where the IG and SS wave height are similar.
Interactions between three IG components have been detected on mild sloping beaches, and have been associated with IG waves pitching forward and breaking \citep{VanDongeren2007,Bakker2015,DeBakker2016}.

Weakly nonlinear energy balances have significantly improved our understanding of nearshore IG wave dynamics but typically break down (i.e., the balance does not close) in the strongly nonlinear inner surfzone \citep[e.g.,][]{Guedes2013,Rijnsdorp2015a,Ruju2012,Mendes2018InfragravityStudy}. Neglected or inaccurately estimated nonlinear terms distort our understanding of the dissipation of wave energy $D_f$ when it is estimated as the residual of the energy balance (Eq. \ref{eq:EB1}). Here, we consider a fully nonlinear wave energy balance for shallow water, where $a/d= \delta=O(1)$ and dispersive effects are cooperatively small $O(\mu) \ll O(\delta)$ (i.e., Ursell numbers $Ur= \delta/\mu^2 \gg 1$).  We derive a frequency-resolved energy balance following the methodology of \citet{Henderson2002} in Section \ref{sec:Theory} based on the nonlinear shallow water equations that are expected to well describe the macro-properties of nonlinear and non-dispersive wave motions. Using simulations with a fully nonlinear and dispersive wave model (SWASH, Section \ref{sec:NM}) of irregular waves propagating over a moderate (1/30) and mild (1/100) planar slope (Section \ref{sec:M-DD}), we use the energy balance to study nearshore IG wave dynamics (Section \ref{sec:Results}). This includes a description of the spatial variations of the nonlinear interactions  contributing to the IG energy balance, and an analysis of the terms responsible for nearshore IG flux losses. In Section \ref{sec:Discussion}, the new NLSWE energy balance is compared with existing theories, the role of IG wave breaking is described,
and limitations of the nondispersive energy balance for SS waves are discussed.  Results are summarized in Section \ref{sec:Conclusions}

\section{Theory}\label{sec:Theory}
\subsection{Bulk energy balance}\label{sec:T-BEB}
We consider shallow water waves in a horizontal ($x,y$) plane with a single-valued free-surface $z=\eta(x,y,t)$, bottom $z=-d(x,y)$, with components of the depth-averaged flow $u_m$, and continuous space and time derivatives.
Assuming a very small characteristic depth over wavelength $\mu$ (or equivalently, assuming negligible vertical accelerations) yields the two-dimensional nonlinear Shallow Water Equations (NLSWE),
\begin{eqnarray}
     \frac{\partial \eta}{\partial t} + \frac{\partial D u_n}{\partial x_n} &=& 0,\label{eq:mass}\\
    \frac{\partial u_m}{\partial t} +  u_n\frac{\partial u_m}{\partial x_n} + g\frac{\partial \eta}{\partial x_m} &=& -\frac{\tau_m}{D}\label{eq:mom}.
\end{eqnarray}
The Einstein summation convention is used, with the instantaneous water depth $D=d+\eta$, gravity acceleration  $g$, and seabed stress $\tau_m$. A momentum balance in conservative form is retrieved by adding Eq. \eqref{eq:mom} $\times D$ and Eq. \eqref{eq:mass} $\times u_m$,
\begin{equation}
\frac{\partial Q_m}{\partial t} + \frac{\partial }{\partial x_n} \left( Du_n u_m + \frac{1}{2}gD^2 \right) = gD\frac{\partial d}{\partial x_m}  - \tau_m,\label{eq:momcons}
\end{equation}
with $Q_m=Du_m$.

An average bulk energy balance follows from multiplying Eq. \eqref{eq:momcons} with $u_m/2$; Eq. \eqref{eq:mom} with $Du_m/2$; Eq. \eqref{eq:mass} with $g\eta$, summing, and taking the expected value,
\begin{equation}
    \frac{\partial E}{\partial t} +  \frac{\partial F_n}{\partial x_n} = S^\tau.
    \label{eq:encons}
\end{equation}
The energy density $E$, energy flux $F$ and a frictional dissipation source $S^\tau$ are defined as,
\begin{eqnarray}
    E &=& \frac{1}{2}g \left\langle\eta^2\right\rangle + \frac{1}{2}\left\langle Du_m u_m \right\rangle\!,\\
    \label{eq:bulkflux}
    F_n &=& \left\langle Q_n \left[   \frac{1}{2}u_m u_m + g\eta \right]\right\rangle,\\
    S^\tau &=& -\left\langle\tau_m u_m\right\rangle\!,\label{eq:bulktaub}
\end{eqnarray}
where $\left\langle \ldots \right\rangle$ denotes a moving average over the fast time scale of the longest considered wave motions. With $S^\tau$ neglected and a smooth solution,  $E$ is conserved to leading order. For stationary one-dimensional conditions, $\frac{\partial E}{\partial t}= \frac{\partial F}{\partial x}=\frac{\partial Q}{\partial x}=0$, the conservation of hydraulic head is recovered,
\begin{equation}
    \frac{u^2}{2} + g\eta = \textrm{constant}.
\end{equation}

Eq. \eqref{eq:momcons} lacks mechanisms for wave breaking (e.g., wave overturning and air-entertainment), and for transferring organized energy to turbulence and heat. With breaking waves, the NLSWE-based energy balance does not close and the balance residual is interpreted as the dissipation rate of organized energy.

\subsection{Frequency resolved energy balance}\label{T-FEB}
A frequency-resolved energy balance details energy exchanges between different frequencies and identifies the frequency bands with the largest residual (dissipation). Following \citet{Henderson2002}, a variable $X(t)$ is Fourier decomposed over frequencies $f$ spaced $\Delta f$ apart,
\begin{equation}
    X(t) = \sum_f \overline{X}^f e^{\mathrm{i} \omega t},
    \label{eq:fourier}
\end{equation}
where $\omega=2\pi f$, $\overline{X}^f(x,y,t)$ denotes the complex amplitude of the frequency decomposition. To ensure real valued functions $(\overline{X}^f)^* = \overline{X}^{-f}$, where $\mathrm{*}$ indicates the complex conjugate. The temporal variation of spectral amplitudes is associated with slow scale changes in mean energy (on scale $T$), whereas the complex exponential term accounts for oscillatory behaviour (with scale $f^{-1}$). The lowest considered frequency $f_0$ in principal separates wave-like dynamics from those associated with "very low frequency" vortical flows that are not included. A meaningful decomposition requires that means change on a much slower time scale $T$ than the longest wave frequencies considered (i.e., $f_0^{-1} T \ll 1$). The assumptions of scale separation and negligible vortical flow in the IG band might be problematic in the strongly nonlinear inner surfzone, but relaxing these assumptions is beyond the present scope.

Frequency balances follow from substitution of Eq. \eqref{eq:fourier} for all time-dependent variables into Eqs. \eqref{eq:mass}-\eqref{eq:momcons}, collecting terms at like frequencies, and multiplication by $e^{-\mathrm{i} \omega t}$:
\begin{eqnarray}
    &\frac{\partial\overline{\eta}^f}{\partial t} + \mathrm{i}\omega \overline{\eta}^f +\frac{\partial \overline{{D u_m}}^f  }{\partial x_m} = 0,  \label{eq:I} \\
    &\frac{\partial \overline{u_m}^f}{\partial t} + \mathrm{i}\overline{u_m}^f +  \overline{ u_n  \frac{\partial u_m}{\partial x_n}}^f + g\frac{\partial \overline{\eta}^f }{\partial x_m} = -\overline{\left(\frac{\tau_m}{D}\right)}^f\!, \label{eq:III} \\
&\frac{\partial \overline{D u_m}^f}{\partial t} + \mathrm{i}\omega \overline{D u_m}^f +   \frac{\partial \overline{D u_n u_m}^f}{\partial x_n} + g  \overline{D \frac{\partial \eta}{\partial x_m} }^f = -\overline{\tau_m}^f\!.
\label{eq:II}
\end{eqnarray}
A balance for the temporal mean of the kinetic energy density follows from combining Eq. \eqref{eq:III} and \eqref{eq:II}\!,
\begin{equation}
   \left( \frac{1}{4} \overline{D u_m}^f \times \mathrm{Eq.} \eqref{eq:III}^* + \mathrm{C.C.} \right) + \left( \frac{1}{4} \left(\overline{u_m}^f\right)^* \times \mathrm{Eq.} \eqref{eq:II} + \mathrm{C.C.} \right)\!.
\end{equation}
Dividing by $\Delta f$, considering the limit $\Delta f \rightarrow 0$, taking expected values, and using the definition of the co-spectrum,
\begin{equation}
  \mathrm{C}_f\!\left(X;Y\right) = \lim_{\Delta f \rightarrow 0} \mathrm{Re}\left\{\left\langle\frac{    \overline{X}^f \left(\overline{Y}^f\right)^* \frac{}{}  }{\Delta f}\right\rangle\right\}\!,
\end{equation}
results in the following kinetic energy balance,
\begin{multline}
  \frac{1}{2} \frac{\partial}{\partial t} \mathrm{C}_f\! \left( Du_m;u_m  \right) + g \; \mathrm{C}_f\! \left( Du_m;\frac{\partial \eta}{\partial x_m}\frac{}{} \right) + \frac{1}{2} \frac{\partial}{\partial x_m} \left[  \mathrm{C}_f\!\left( D u_n u_m;u_m \frac{}{} \right) \right] \\
  = \frac{1}{2} \mathrm{C}_f\!\left( D u_m u_n; \frac{\partial u_m}{\partial x_n} \right) - \frac{1}{2} \mathrm{C}_f\!\left( D u_m; u_n \frac{\partial u_m}{\partial x_n} \right) \\
  + \frac{1}{2} g \; \mathrm{C}_f\! \left( D u_m; \frac{\partial \eta }{\partial x_m} \right) - \frac{1}{2} g \; \mathrm{C}_f\! \left( u_m; D \frac{\partial\eta}{\partial x_m} \right) - \frac{1}{2} \mathrm{C}_f\! \left( D u_m; \frac{\tau_m}{D} \right) - \frac{1}{2} \mathrm{C}_f\! \left(u_m;\tau_m\frac{}{}\right)\!.
\end{multline}
A potential energy balance is derived by
multiplying  Eq. \eqref{eq:I} with $\frac{1}{2}g \overline{\eta}^f$, followed by steps similar to the kinetic balance,
\begin{equation}
  \frac{1}{2} g \frac{\partial}{\partial t} \mathrm{C}_f\! \left( \eta; \eta \frac{}{} \right) +  g \; \mathrm{C}_f\! \left( \eta; \frac{\partial Du_m}{\partial x_m} \right) = 0.
\end{equation}
Combining the potential and kinetic energy balances yields a frequency balance for the total organised energy,
\begin{equation}
  \frac{\partial E_f}{\partial t} + \frac{\partial }{\partial x_n}\left( F^\mathrm{L}_{n,f}+F^{\mathrm{NL}}_{n,f}\right) = S^{\mathrm{NL}}_f + S^{\tau}_f,
  \label{eq:enbalance}
\end{equation}
where the flux $F_{n,f}$ is decomposed into a (quasi-)linear (superscript L) and nonlinear (superscript NL) contribution (i.e., $F_{n,f} = F_n^{\mathrm{L}} + F_n^{\mathrm{NL}}$). In a linear approximation the correlation between $u$ and $\eta$ is the only contribution to the flux. For this reason we will consider $F_n^{\mathrm{L}}$ the `linear' part of the flux. However, both $u$ and $\eta$ contain nonlinear corrections, and the correlation does not exactly equal the $E c_{g}$ fully linear approximation. The other frequency dependent terms are given by,
\begin{eqnarray}
  E_f &=& \frac{1}{2}g \mathrm{C}_f\! \left( \eta; \eta \frac{}{} \right) + \frac{1}{2} \mathrm{C}_f\! \left( D u_m; u_m \frac{}{} \right)\!,\label{eq:EB-E}\\
  F^{\mathrm{L}}_{n,f} &=& g \mathrm{C}_f\! \left( du_n ; \eta \frac{}{} \right)\!,\label{eq:EB-F-L}\\
  F^{\mathrm{NL}}_{n,f} &=& g \mathrm{C}_f\! \left( \eta u_n ; \eta \frac{}{} \right) + \frac{1}{2} \mathrm{C}_f\! \left( D u_n u_m; u_m \frac{}{} \right)\!,\label{eq:EB-F-NL}\\
  S_f^{\textrm{NL}} &=& \frac{1}{2} \left[ \mathrm{C}_f\!\left( D u_m u_n; \frac{\partial u_m}{\partial x_n} \right) - \mathrm{C}_f\!\left( D u_m; u_n \frac{\partial u_m}{\partial x_n} \right) \right]\label{eq:EB-Snl}\\
  &+& \frac{1}{2} g \left[ \mathrm{C}_f\! \left( \eta u_m; \frac{\partial\eta}{\partial x_m} \right) - \mathrm{C}_f\! \left( u_m; \eta \frac{\partial\eta}{\partial x_m} \right) \right]\!,\nonumber\\
  S_f^{\tau} &=& - \frac{1}{2} \mathrm{C}_f\! \left( D u_m; \frac{\tau_m}{D} \right) - \frac{1}{2} \mathrm{C}_f\! \left(u_m;\tau_m\frac{}{}\right)\label{eq:EB-tau}\!.
\end{eqnarray}
Note that cross-shore variation of the still water depth is included at lowest order in Eq. \eqref{eq:enbalance} through gradients of $F^{\mathrm{L}}_{n,f}$ (Eq. \ref{eq:EB-F-L})

This frequency resolved energy balance is the central result of this work. Apart from the change to a spectral formulation, the balance derivation  mirrors that of the bulk balance. The bulk energy, energy flux and frictional dissipation terms (Eqs. \ref{eq:encons}-\ref{eq:bulktaub}) all have clear frequency-domain counterparts (Eqs. \ref{eq:EB-E}-\ref{eq:EB-tau}). The bulk and frequency-resolved formulations are internally consistent because
\begin{equation}
    \int C_f \left( X; Y \right) \,\mathrm{d}f =  \left\langle XY\right\rangle\!,
\end{equation}
and
\begin{equation}
    \int \mathrm{C}_f \!\left( X; YZ \right) \mathrm{d}f = \int \mathrm{C}_f\! \left( XY;Z \right) \mathrm{d}f = \left\langle XYZ\right\rangle\!.
\end{equation}
Bulk expressions for $E$, $F_n$ and $S^\tau$ are recovered when $E_f$, $F_{n,f}$ and $S^\tau$ are integrated over all frequencies. Further, integrated contributions from the nonlinear interaction term $S^{\textrm{NL}}_f$ vanish,  consistent with the expectation that nonlinear interactions redistribute energy across frequencies but conserve total energy.

This spectral energy balance accounts for all mechanisms potentially transferring energy between SS and IG waves, including the nonlinear shoaling of bound waves \citep[e.g.,][]{List1992,Janssen2003,Battjes2004}, excitation of free IG waves over a sloping bed \citep[e.g.,][]{Mei1984,Nielsen2010,Contardo2021FreeBreaking,Liao2021AnConditions} and breakpoint generation \citep{Symonds1982}. However, within the spectral framework all these mechanisms are represented as either contributions to nonlinear flux gradients or nonlinear interactions, so that identifying the dominant mechanism that drives the interaction is in general not possible. Here, we therefore quantify the spectral energy flow but otherwise will not attempt to interpret these in the context of mechanisms proposed in the literature.

\section{Methodology}\label{sec:NM}

\subsection{Numerical model}
The highly detailed flow variables required to evaluate the frequency-resolved energy balance (Eq. \ref{eq:enbalance}) in the surfzone are not readily available from field or laboratory experiments. Instead, we use the nonlinear and fully dispersive wave model SWASH to obtain the required variables. SWASH is a multi-layer non-hydrostatic wave-flow model, and essentially a direct numerical implementation of the Reynolds Averaged Navier Stokes (RANS) equations \citep{Zijlema2011}. Previous studies have shown that SWASH accurately describes the nonlinear transformation and breaking of SS waves \citep[e.g.,][]{Smit2013,Smit2014}, the IG wave dynamics \citep[e.g.,][]{Rijnsdorp2014,DeBakker2016,Fiedler2019TheModeling}, and run-up oscillations at the beach \citep[e.g.,][]{Ruju2014,Lerma2017}.

\subsection{Model set-up}\label{sec:NM-setup}
The energy balance is studied for SS waves normally incident on mild (1/100) and moderately (1/30) sloping beaches. Irregular waves were generated in 15 m depth using a weakly reflective wavemaker to avoid re-reflection at IG frequencies. The irregular wave field had a JONSWAP spectral shape with significant wave height $H_{m0}=3$ m and peak period $T_p=10$ s. The wavemaker signal was based on weakly nonlinear theory \citep{Hasselmann1962} to suppress generation of free IG waves \citep[see][for further details]{Rijnsdorp2014}.

A horizontal resolution of 1 m was used, which is approximately $1/70$ of the peak wave length at the edge of the surfzone, in combination with a time-step of 0.025 s (corresponding to a Courant number $CFL\approx0.3$). A fine vertical resolution (10 vertical layers) was used to ensure that the kinematic condition for the onset of wave breaking (i.e., the orbital velocity exceeds the wave celerity) is captured accurately and wave breaking occurs at the correct location \citep{Smit2013}.
The tangential bottom stress $\tau_b$ was estimated using the law of the wall with roughness height $4\times10^{-4}$ m, a representative value for smooth concrete \citep[e.g.,][]{Chow1959}. The $k-\epsilon$ turbulence model accounts for vertical mixing due to shear in the water column but does not describe overturning waves, air entrainment, breaking generated turbulence, or the transfer of organized energy into turbulence and heat. In NLSWE-based models like SWASH, waves steepen until  a jump-discontinuity (or shock) develops that represents a broken wave. SWASH uses the momentum-conserving (shock-capturing) numerical scheme of \citet{Stelling2003} to solve the momentum equations (Eq. \ref{eq:mom}) in their conservative form (Eq. \ref{eq:momcons}). As a result, jumps-discontinuities dissipate total energy in accordance with a hydraulic jump. Such shock-capturing numerical schemes allows NLSWE-based wave models to simulate the bulk dissipation of a breaking wave without accounting for wave-breaking generated turbulence \citep[e.g.,][]{Tissier2012,Smit2013}.

The model depth-averaged horizontal velocities and surface elevation were sampled with a horizontal resolution of 1 m and a temporal resolution of 20 Hz, sufficient to estimate accurately the terms of the energy balance at IG frequencies (Appendix \ref{sec:A-D}). The run-up signal was defined as the location where the instantaneous water depth ($d+\eta$) was smaller than 5 cm.  Variables were output for 240 min ($>1000$ peak wave periods) after a spin-up time of 10 min.
\subsection{Data analysis}\label{sec:M-DA}

Time series of velocity or sea surface elevation $X$ are separated with,
\begin{equation}
    X=X_\mathrm{vlf}+X_\mathrm{ig}+X_\mathrm{ss},
\end{equation}
where $X_\mathrm{ig}$ represents the IG frequency band (defined as $\frac{1}{20}f_p<f\leq0.5f_p$, with $f_p$ the peak period at the numerical wavemaker) and $X_\mathrm{ss}$ corresponds to the SS frequency band (defined as $f > 0.5 f_p$). Frequencies below the IG band are considered part of the very-low frequency (VLF) band ($f\leq\frac{1}{20}f_p$). In the following, we use labels with a normal font (in subscripts, for readability) and capital font (in-line text) to distinguish between variables belonging to the VLF, IG and SS frequency band and from combinations thereof (separated by a comma in subscripts and by an em dash for in-line text).

Spectral moments were computed by integrating surface elevation and runup spectra over their respective frequency bands ($m_n=\int E f^n \rm{d}f$). The significant wave height $H_{m0}$ and runup height $R_{m0}$ were computed as $4 \sqrt{m_0}$ with $m_0$ the zeroth-order moment  of the surface elevation or runup spectra. The mean wave period was computed as $T_{m01}=\frac{m0}{m1}$. To quantify the bulk wave nonlinearity, the Ursell parameter $\text{Ur}=\delta/\mu^2$ was computed based on the significant SS wave height ($\delta={a}/({d+\overline{\eta}})$, with the wave amplitude $a=\frac{1}{2}H_{m0,\mathrm{SS}}$) and the mean wave period ($\mu= k d$, with wave number $k$ computed based on $T_{m01,\mathrm{SS}}$). Similarly, we also estimated the nonlinearity of the IG waves by estimating the wave amplitude as $a=\frac{1}{2}H_{m0,\mathrm{IG}}$ and the wave number based on the mean IG wave period $T_{m01,\mathrm{IG}}$.

The frequency dependent NLSWE energy balance (Eq. \ref{eq:enbalance}) was evaluated based on the modelled (depth-averaged) flow and free surface variables. To compare the NLSWE energy balance with existing theories, we evaluated the new theory and the energy balance terms of \citet{Henderson2002,Henderson2006} and \citet{Herbers1997} in a consistent manner. All spectra and cross-spectra in this work were computed with $50\%$ overlapping Hanning windows and a segment length of 8000 samples, resulting in a frequency resolution $\Delta f=0.0025$ Hz.

To quantify the contribution from IG and SS frequencies to the NLSWE energy balance, terms of the frequency dependent energy balance were integrated over both the SS and IG frequency bands. Furthermore, we quantified how different nonlinear contributions affected the nonlinear flux (Eq. \ref{eq:EB-F-NL}) and nonlinear interaction term (Eq. \ref{eq:EB-Snl}) (see Appendix \ref{sec:A-A}). The time-domain analysis \citet{Fiedler2019TheModeling} avoids the cumbersome bookkeeping of traditional estimates based on bispectral integrations \citep[e.g.,][]{Bakker2015}. With this approach, the frequency-dependent nonlinear flux term $F^\mathrm{NL}_f$ and nonlinear-interaction term $S_f^\mathrm{NL}$ were decomposed as,
\begin{eqnarray}
  &F^\mathrm{NL}_f&=F^\mathrm{NL}_{f,\mathrm{ig,ig,ig}}+F^\mathrm{NL}_{f,\mathrm{ig,ig,ss}}+F^\mathrm{NL}_{f,\mathrm{ig,ss,ss}}+F^\mathrm{NL}_{f,\mathrm{ss,ss,ss}}+F^\mathrm{NL}_{f,\mathrm{vlf}},\\
  &S_f^\mathrm{NL}&=S^\mathrm{NL}_{f,\mathrm{ig,ig,ig}}+S^\mathrm{NL}_{f,\mathrm{ig,ig,ss}}+S^\mathrm{NL}_{f,\mathrm{ig,ss,ss}}+S^\mathrm{NL}_{f,\mathrm{ss,ss,ss}}+S^\mathrm{NL}_{f,\mathrm{vlf}}.
\end{eqnarray}
In these equations, terms with subscript $[...]_\mathrm{ig,ig,ig}$ and $[...]_{\mathrm{ss,ss,ss}}$ represent correlations between three IG and three SS wave components, respectively. Terms with $[...]_{\mathrm{ig,ig,ss}}$ represent correlations between two IG and one SS wave component, terms with $[...]_{\mathrm{ig,ss,ss}}$ represent correlations between a single IG and two SS wave components, and the terms with subscript $[...]_\mathrm{vlf}$ account for correlations between three wave components of which at least one is a VLF component.

Integrating the above frequency dependent $F^\mathrm{NL}_f$ and $S_f^\mathrm{NL}$ over the IG frequencies results in the decomposed bulk IG flux and bulk IG nonlinear interaction term,
\begin{eqnarray}
  &F^\mathrm{NL}_\mathrm{IG}&=F^\mathrm{NL}_\mathrm{IG,ig,ig,ig}+F^\mathrm{NL}_\mathrm{IG,ig,ig,ss}+F^\mathrm{NL}_\mathrm{IG,ig,ss,ss}+F^\mathrm{NL}_\mathrm{IG,vlf},\\
  &S_\mathrm{IG}^\mathrm{NL}&=S^\mathrm{NL}_\mathrm{IG^\pm,ig,ig,ig}+S^\mathrm{NL}_\mathrm{IG,ig,ig,ss}+S^\mathrm{NL}_\mathrm{IG,ig,ss,ss}+S^\mathrm{NL}_\mathrm{IG,vlf}.
\end{eqnarray}
In these decomposed terms, correlations between three SS components are not included as they do not contribute to IG fluxes. Furthermore, $S^\mathrm{NL}_\mathrm{IG,ig,ig,ig}$ integrates to zero over the IG band as interactions between three IG components redistribute energy but do not account for a net energy transfer to or away from IG frequencies. In order to quantify the energy flow within the IG band by $S^\mathrm{NL}_\mathrm{IG,ig,ig,ig}$, we integrated $S^\mathrm{NL}_{f,\mathrm{ig,ig,ig}}$ over the IG band by considering either positive or negative interactions (denoted as $S^\mathrm{NL}_\mathrm{IG^\pm,ig,ig,ig}=S^\mathrm{NL}_\mathrm{IG^+,ig,ig,ig}+S^\mathrm{NL}_\mathrm{IG^-,ig,ig,ig}$).

To evaluate the biphase $\beta$ for the individual IG and SS band and the correlations between both bands, we computed the asymmetry ($\mathrm{As}$) and skewness ($\mathrm{Sk}$) in the time-domain for all possible triads (excluding the VLF motions) following the approach of \citet{Fiedler2019TheModeling}. Subsequently, we computed the biphase $\beta$ and bicoherence for each combination of triads. For example, the biphase of triads between three SS components was computed as,
\begin{equation}
    \beta_\mathrm{ss,ss,ss}=\tan^{-1}\left(\frac{\mathrm{As}_\mathrm{ss,ss,ss}}{\mathrm{Sk}_\mathrm{ss,ss,ss}}\right)\!,
\end{equation}
with a corresponding bicoherence of,
\begin{equation}
  \sqrt{\mathrm{As}_\mathrm{ss,ss,ss}^2+\mathrm{Sk}_\mathrm{ss,ss,ss}^2}.
\end{equation}

\section{Bulk wave evolution}\label{sec:M-DD}
\begin{figure}
\captionsetup{singlelinecheck = false, justification=justified}
    \centering
    \includegraphics[scale=0.8]{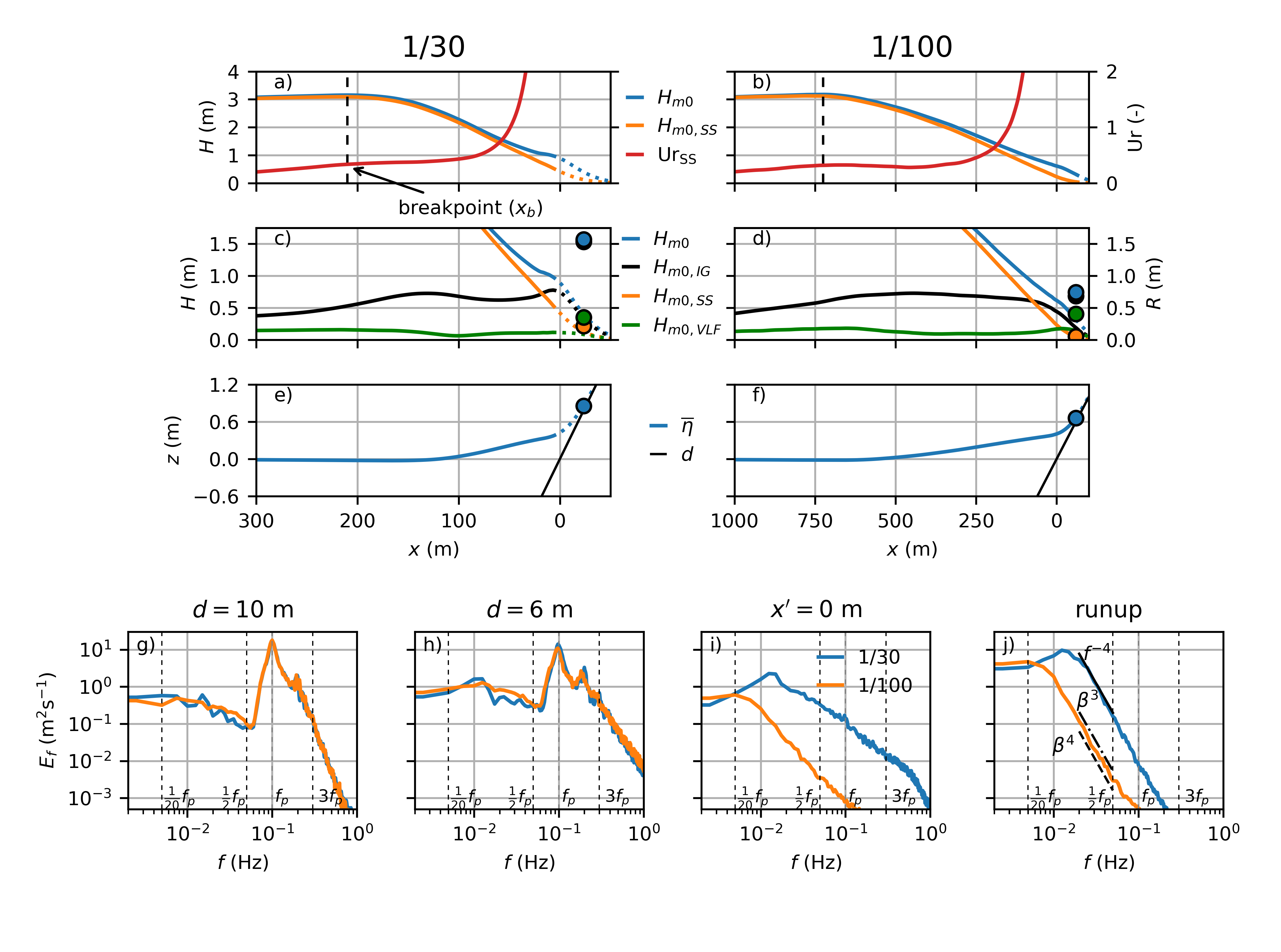}
     \caption{
    Overview of bulk wave evolution. (a-f) total, SS, IG and VLF wave height $H$ and runup $R$, setup $\overline{\eta}$ and still water depth $d$ (see legends) versus cross-shore location $x$. The still water depth is zero at $x=0$ , and $x>0$ offshore. Filled circles in (c,d) indicate significant runup (with values relative to the right axis). In (a-f), cells that are always or intermittently wet are shown with solid and dashed curves, respectively. Dashed vertical line indicates the outer edge of the surfzone, where the total wave height is largest). (g-j) power spectra of surface elevation and runup on 1/30 and 1/100 slopes (blue and orange curves, respectively) at (g) $d=10$ m,  (h) $d=6$ m (approximate SS break point), (i) $x^\prime=0$ (the last cell that is always wet), and (j) of the runup signal. In (j), black lines indicate $f^{-4}$ slopes separated by distance $\beta^{3}$ (dash-dot) and $\beta^4$ (dashed).
    }
    \label{fig:bulk}
\end{figure}

\begin{figure}
\captionsetup{singlelinecheck = false, justification=justified}
     \centering
     \includegraphics{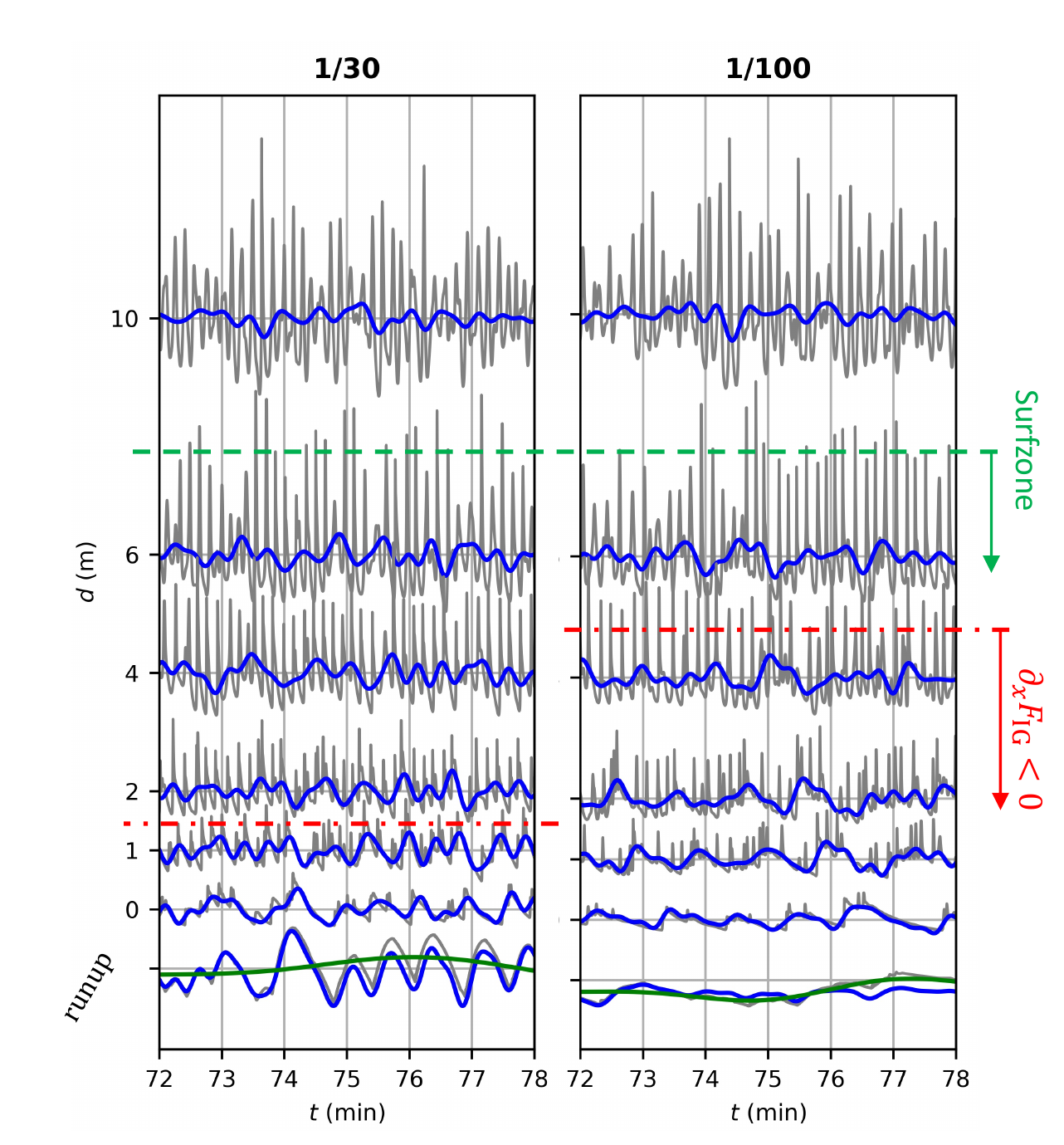}
     \caption{
     Time series of the total surface elevation (thin grey) and the band-passed IG surface elevation signal (blue) from $d=10$ m depth (top) to $d=0$ m depth (bottom) on slopes of (left) 1/30 and (right) 1/100.  At the bottom of both panels the total (thin grey), band-passed IG (blue) and VLF (green) runup signal is shown. For improved visualisation, the IG signal of the surface elevation and runup is translated vertically to oscillate around the mean of the respective total signal. The surfzone, where SS wave energy decreases, starts around $d=6-7$ m on both slopes (indicated by the green dashed line). The IG energy begins to decrease (flux gradient $\partial_x F_\mathrm{IG}<0$) in about 1.5 m and 4 m depth on the 1/30 and 1/100 slope, respectively (indicated by the red dashed line).
     }
     \label{fig:Timestacks}
\end{figure}

The cross-shore evolution of the wave height ($H_\mathrm{IG}$, $H_\mathrm{SS}$) and wave spectrum
illustrate the well known transition from SS dominance offshore (depth $d=10$ m) to IG dominance near the shoreline and in runup (Fig. \ref{fig:bulk}).
On both slopes, the wave height $H_{m0}$ increased slightly during shoaling, followed by breaking at $d\approx6$ m (with the breakpoint $x_b$ located at $x{_b}\approx200$ and $700$ m for the 1/30 and 1/100 slope, respectively) (Fig. \ref{fig:bulk}a,b). During shoaling, energy at the higher harmonics increased gradually, associated with nonlinear wave-interactions between SS components (e.g., $2f_p$ in Fig. \ref{fig:bulk}g-h). This spectral signature is consistent with gradual steepening of SS waves as they shoal and break (Fig. \ref{fig:Timestacks}).

On both slopes, the IG wave height $H_{m0,\mathrm{IG}}$ was smaller than the SS wave height $H_{m0,\mathrm{SS}}$ seaward of and within most of the surfzone. Seaward of the surfzone, IG waves were out of phase with the forcing SS wave groups (Fig. \ref{fig:Timestacks}), consistent with bound IG waves \citep[e.g.,][]{LHS1962}. On the 1/100 slope, $H_{m0,\mathrm{IG}}> H_{m0,\mathrm{SS}}$ for $x<100$ m, whereas on the 1/30 slope, IG motions were largest only near the shoreline ($x<20$ m) (Fig. \ref{fig:bulk}c,d,i). On both slopes, motions at IG frequencies dominated the runup signal (Fig. \ref{fig:bulk} and Fig. \ref{fig:Timestacks}). The filtered time signals indicate that IG waves gradually pitch forward on the 1/100 slope to a bore-like front with small SS waves riding on their crest (Fig. \ref{fig:Timestacks}).

On both slopes,  IG energy levels of the runup spectra were much larger compared to the surface elevation spectra at $x^\prime=0$ (the shallowest location where a cell is always wet), whereas runup energy levels at SS frequencies were lower compared to SS spectral energy levels at $x^\prime=0$ (Fig. \ref{fig:bulk}i,j).
The runup energy levels roll-off as $f^{-4}$ at SS frequencies, and the roll-off (often associated with saturation) extends well into the IG band (Fig. \ref{fig:bulk}j). For the same incident waves, the runup energy levels depend on the beach slope $\beta$ and are separated by a distance proportional to $\beta^{-l}$ where $l=3$ and $4$ for the present slopes of 1/30 and 1/100, respectively.

During shoaling ($x>x_{b}$), the bulk SS Ursell parameter $\text{Ur}_\mathrm{SS}\approx 0.5$. Here, dispersive and nonlinear effects are equally important in the SS band, indicating that the classical Bousinesq scaling holds. The present energy balance based on the NLSWE is nondispersive and does not fully explain the physics of shoaling SS waves. However, the NLSWE energy balance is potentially accurate in the surfzone and especially the inner surfzone, where nonlinearity dominates over dispersion ($\text{Ur}_\mathrm{SS}>1$). The IG Ursell parameter $\text{Ur}_\mathrm{IG}>1$ for $d<10$ m (not shown), and the NLSWE energy balance is expected to explain the IG wave dynamics throughout the domain.

\section{Energy balance}\label{sec:Results}

\subsection{Sea Swell (SS) balance }
\begin{figure}
\captionsetup{singlelinecheck = false, justification=justified}
    \centering
     \includegraphics[scale=0.9]{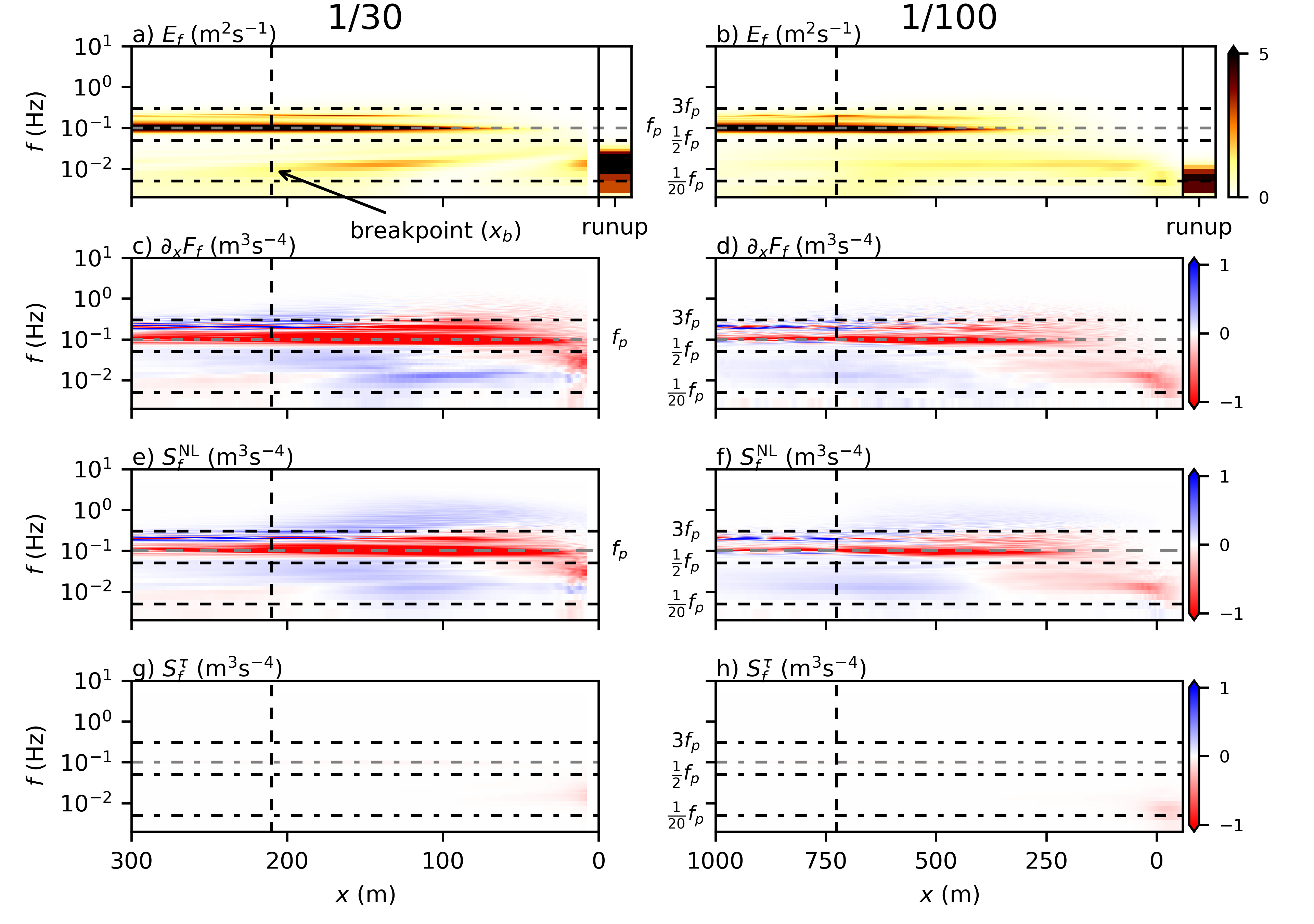}
    \caption{Cross-shore variation of the surface elevation and runup spectra (a,b) and frequency-dependent NLSWE energy balance terms (c-h) on a 1/30 (left) and 1/100 slope (right).
     The horizontal gray dotted line indicates the offshore peak frequency $f_p$ and black dashed lines indicate the limits of the infragavity ($\frac{1}{20}f_p<f\leq f_p/2$), primary SS ($\frac{1}{2}f_p<f\leq 3f_p$), and superharmonic ($f>3f_p$) frequency bands. The dashed vertical line indicates the breakpoint $x_b$ ($d\approx 6$ m).
     }
    \label{fig:Example_freq}
\end{figure}

\begin{figure}
\captionsetup{singlelinecheck = false, justification=justified}
    \centering
    \includegraphics[scale=0.9]{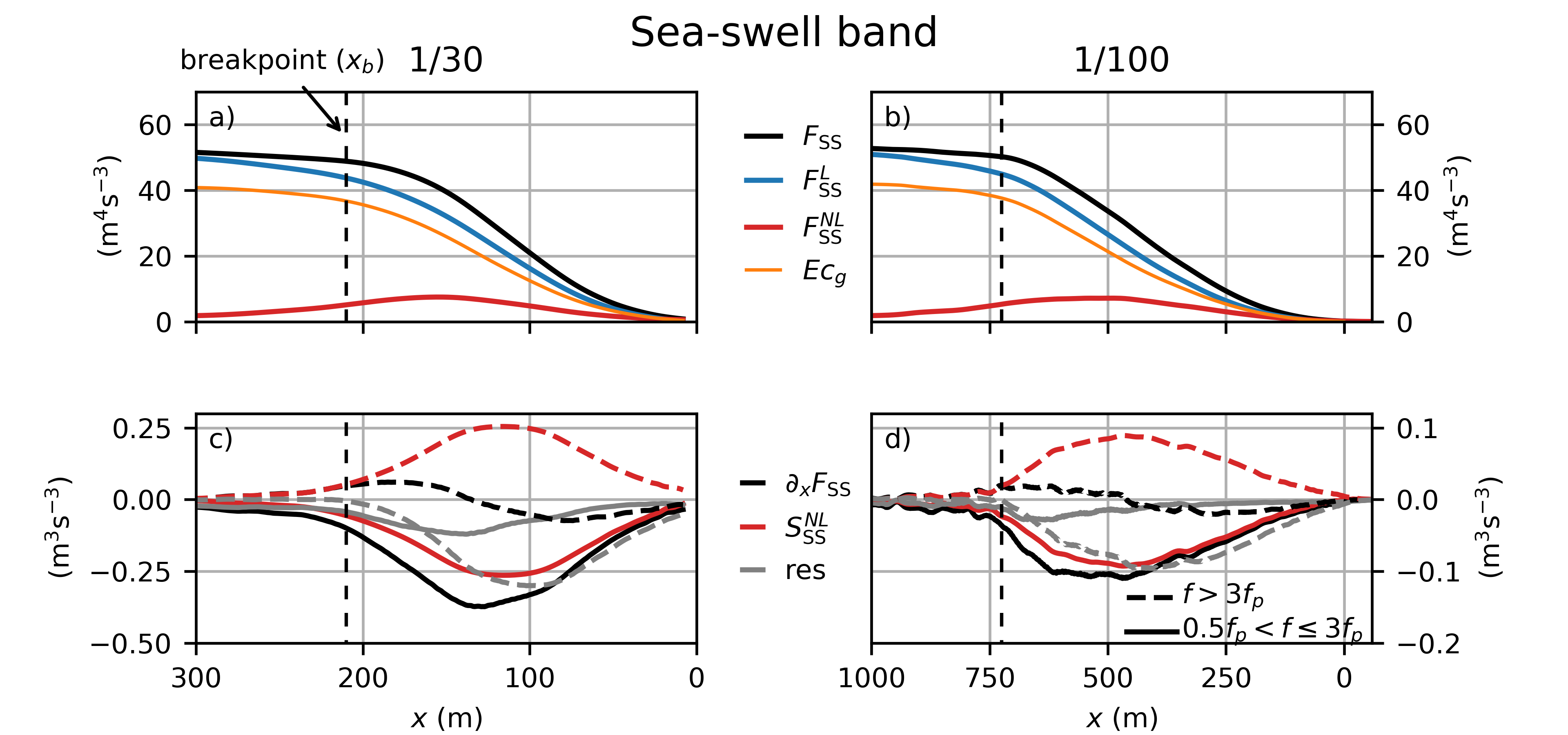}
    \caption{Cross-shore variation of the bulk fluxes (a,b) and non-zero bulk NLSWE energy balance terms (c,d) of the SS band on slopes of (left) 1/30 and (right) 1/100. The balance terms in (c,d) are further decomposed into primary ($\frac{1}{2}f_p<f\leq 3f_p$) and superharmonic ($f>3f_p$) SS frequencies. The dashed vertical line indicates the breakpoint $x_b$ ($d\approx 6$ m).}
    \label{fig:Example_SS}
\end{figure}

The cross-shore variation of the (frequency dependent) NLSWE energy balance terms shows that seaward of the surfzone the energy flux increases ($\partial_x F_f>0$) at harmonics $2f_p$ and $3f_p$ and decreases at $f_p$ (blue and red shading in Fig.\ref{fig:Example_freq}c-d for $x>x_{b}$, respectively). These flux gradients are largely balanced by nonlinear interactions $S^\mathrm{NL}_f$ with negligible bottom stress $S^\tau_f$ (Fig. \ref{fig:Example_freq}e-h). Integrating the energy balance terms over the primary SS ($0.5 f_p < f \leq 3 f_p$) and over the super-harmonic frequencies ($f>3f_p$) shows that the energy balance approximately closes seaward of the surfzone ($x>x_b$) on both slopes (grey curves in Fig. \ref{fig:Example_SS}c,d).

Within the surfzone ($x<x_b$), nonlinear interactions strengthen and transfer energy towards increasingly high super-harmonic frequencies ($f>3f_p$, Fig. \ref{fig:Example_freq}e-f). The energy flux $F_f$, however, did not increase at the super-harmonic frequencies (Fig. \ref{fig:Example_freq}c-d). Lacking significant dissipation from bottom stress (Fig. \ref{fig:Example_freq}g-h), the super-harmonic residual ($f>3f_p$) was negative and relatively large in the surfzone (grey-dashed in Fig. \ref{fig:Example_SS}c,d). The residual was smaller in the more energetic primary band (grey-solid in Fig. \ref{fig:Example_SS}c,d). The transfer of SS energy to superharmonic frequencies with breaking waves is generally consistent with previous studies based on weakly nonlinear but dispersive balances \citep[e.g.,][]{Herbers2000,Smit2014}.

\subsection{ Infragravity (IG) balance}

On both slopes nonlinear interactions increase the energy flux at the IG frequencies in the shoaling region and much of the surfzone (pale blue, $S_f^\mathrm{NL}>0$ and $\partial_x F_f>0$  in Fig. \ref{fig:Example_freq}c-f). Deeper inside the surfzone, nonlinear interactions changed sign, resulting in $\partial_x F_f<0$ at higher IG frequencies. On the 1/30 slope, lower IG frequencies continued to receive energy up to close to the shoreline. In contrast, on the 1/100 slope, $\partial_x F_f$ gradually changes sign over the whole IG band and all IG frequencies lose energy for $x<300$ m.  IG dissipation from bottom friction $S_f^{\tau}$ was small except for the lowest frequencies in very shallow water (Fig. \ref{fig:Example_freq}g-h).

\begin{figure}
\captionsetup{singlelinecheck = false, justification=justified}
    \centering
    \includegraphics[scale=0.9]{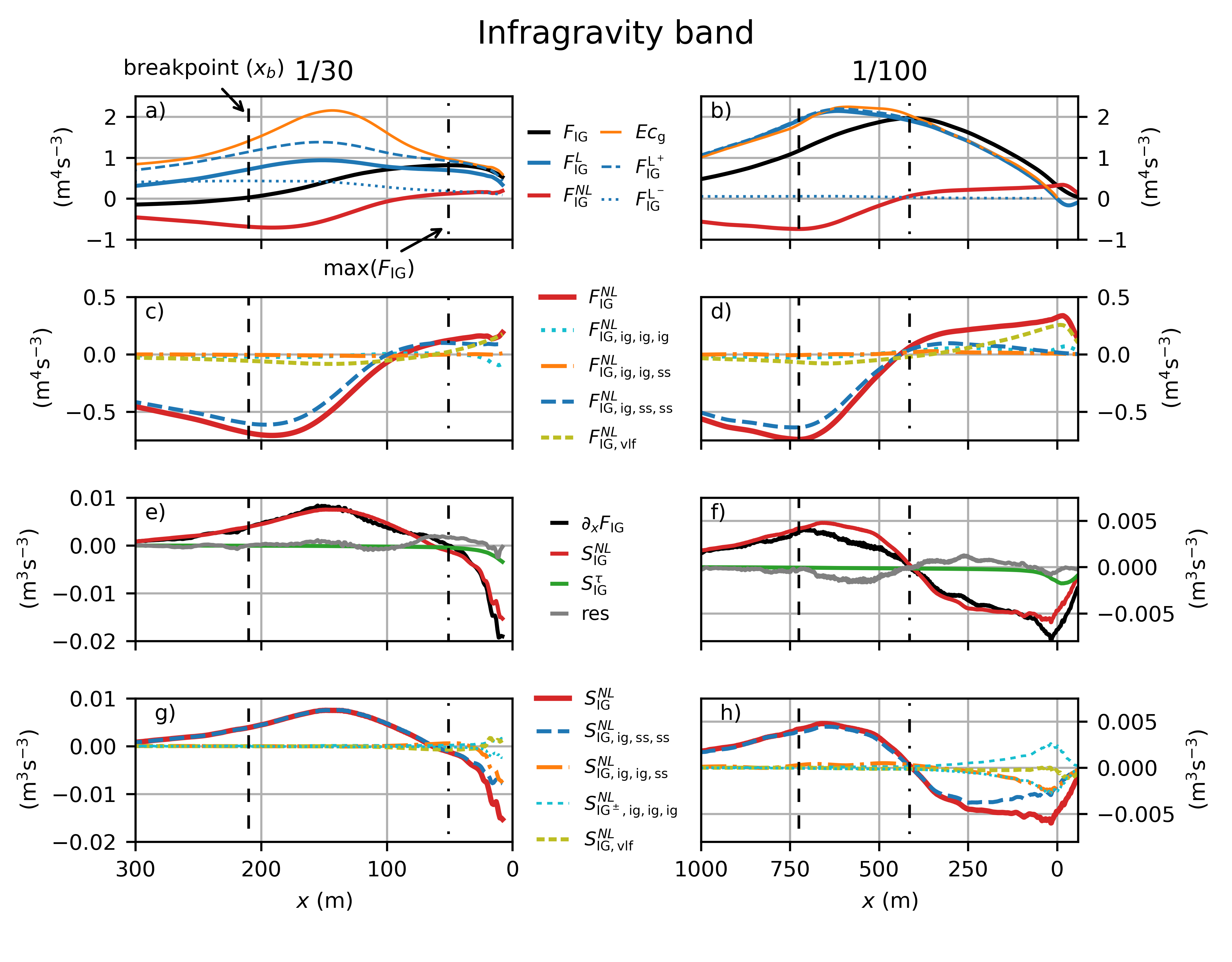}
   \caption{Cross-shore variation of the (decomposed) bulk IG fluxes (a-d), the bulk IG energy balance terms (e,f) and the (decomposed) bulk nonlinear interaction terms (g,h) on a 1/30 (left panels) and 1/100 slope (right panels). The shoreline
   is at $x=0.$
   In (a,b), the decomposition of the linear IG flux $F^\mathrm{L}_\mathrm{IG}$ into shoreward and seaward components ($F^\mathrm{L^+}_\mathrm{IG}$ and $F^\mathrm{L^-}_\mathrm{IG}$, respectively) is not shown when the WKB assumption is violated ($\frac{\omega^2 d}{g}<10~h_x^2$, with $\omega$ from $T_{m01,\mathrm{IG}}$). Vertical lines indicate the location of (dashed) the seaward surfzone edge and (dash-dotted) where the IG flux gradient $\partial_x F_\mathrm{IG}$ changes sign.}
   \label{fig:Example_IG}
\end{figure}

On the 1/100 slope, bulk IG fluxes $F_\mathrm{IG}$ were simplified by the relatively low amplitude of seaward going IG waves ($F_\mathrm{IG}^{\mathrm{L}^+}>>F_\mathrm{IG}^{\mathrm{L}^-}$,  Fig. \ref{fig:Example_IG}b). At $d=10$ m, the relative radiation coefficient
 $R^2_\mathrm{IG}$ = $F_\mathrm{IG}^{\mathrm{L}^-}/F_\mathrm{IG}^{\mathrm{L}^+} \approx 0.05 $ . With this negligible seaward IG radiation,
$F_\mathrm{IG}^{\mathrm{L}} \approx F_\mathrm{IG}^{\mathrm{L}^+} \approx E c_g$ as assumed by \citet{DeBakker2016} and others. On the 1/30 slope,  $R^2_\mathrm{IG} \approx 0.6$ at $d=10$ m, and $E c_g$ is inaccurate (Fig. \ref{fig:Example_IG}a). The accuracy of the present linear estimates of seaward and shoreward IG fluxes for these nonlinear waves is unknown.

On both slopes, the nonlinear contribution  $F_\mathrm{IG}^{\mathrm{NL}}$ to the total IG flux ($F_{IG}$) is significant (Fig. \ref{fig:Example_IG}a-b). In the shoaling region and a large part of the surfzone, $F_\mathrm{IG}^\mathrm{NL}<0$ due to a negative correlation between IG waves and SS components, as occurs with bound IG waves that are $180^o$ out of phase with wave group forcing. In the surfzone, the phase drifts away from $180^o$ \citep[e.g.,][]{Janssen2003}. The cross-shore location of the maximum $F_\mathrm{IG}$ (where  $\partial_x F_\mathrm{IG}$ changed sign) is around $x=450$ m ($\approx0.65 x_b$) on the 1/100 and $x=50$ m ($\approx0.25x_{b}$) on the 1/30 slope (Fig. \ref{fig:Example_IG}a-b). Further shoreward in the inner surfzone, IG waves are losing energy on both slopes.
These IG energy losses ($\partial_x F_\mathrm{IG}<0$) were largely explained by $S^\mathrm{NL}_\mathrm{IG}$ and a small bottom stress $S^\mathrm{\tau}_\mathrm{IG}$. The IG residual is relatively small everywhere (grey lines in Fig. \ref{fig:Example_IG}e-f).

\subsection{Contributors to the flux and nonlinear interactions at infragravity frequencies}\label{sec:Results_Contr}

To quantify the contribution from different triads to the nonlinear interactions, we decomposed $S^\mathrm{NL}_\mathrm{IG}$ into correlations between different combinations of the SS wave signal, the IG wave signal, and the remaining VLF signal (see Section \ref{sec:M-DA} and Appendix \ref{sec:A-A}). Some previous studies  estimated the total flux from the linear flux alone \citep[e.g.,][]{Thomson2006,Bakker2015}, which can be problematic in the nearshore \citep[e.g.,][]{Henderson2006}.
In the present simulations, the linear flux $F_\mathrm{IG}^\mathrm{L}$ under-predicted the total flux $F_\mathrm{IG}$ in the inner surfzone shoreward of the location where $F_\mathrm{IG}$ peaked, and over-predicted $F_\mathrm{IG}$ seaward of the location where $F_\mathrm{IG}$ peaked (compare black with blue curves, Fig. \ref{fig:Example_IG}a,b).
Seaward of the breakpoint, the dominant contributor to $F_\mathrm{IG}^\mathrm{NL}$ is due to interactions between one IG and two SS wave components $F^\mathrm{NL}_\mathrm{IG,ig,ss,ss}$. This is consistent with numerous previous studies \citep[e.g.,][and many others]{LHS1962,Henderson2006,Ruju2012,Mendes2018InfragravityStudy}.

The interactions between one IG and two SS waves, associated with the forcing of IG waves by the SS waves \citep[e.g.,][]{LHS1962,Hasselmann1962}, also (nearly) completely explained $S_\mathrm{IG}^\mathrm{NL}$ (Fig. \ref{fig:Example_IG}g-h). Within the surfzone, $S^\mathrm{NL}_\mathrm{IG,ig,ss,ss}$ did not explain all the negative work by $S_\mathrm{IG}^\mathrm{NL}$, and additional interactions are required to close the IG energy balance. Nonlinear interactions with at least one VLF component ($S^\mathrm{NL}_\mathrm{IG,vlf}$) did not provide a significant contribution (Fig. \ref{fig:Example_IG}g-h). Instead, the inclusion of interactions between two IG and one SS component $S^\mathrm{NL}_\mathrm{IG,ig,ig,ss}$ was required to accurately estimate $S_\mathrm{IG}^\mathrm{NL}$ inside the surfzone. Separating $S^\mathrm{IG}_\mathrm{ig,ig,ig}$ (which integrated to zero over the IG band) into positive and negative contributions ($S^\mathrm{IG^+}_\mathrm{ig,ig,ig}$ and $S^\mathrm{IG^-}_\mathrm{ig,ig,ig}$, respectively) quantifies the energy flow within the IG band due to triad interactions among IG waves. For both slopes, $S^\mathrm{IG^+}_\mathrm{ig,ig,ig}$ and $S^\mathrm{IG^-}_\mathrm{ig,ig,ig}$ became non-zero shoreward of the location where $F_\mathrm{IG}$ peaked. Their contribution was largest for the mild-slope (Fig. \ref{fig:Example_IG}h), where they resulted in a flow of energy from lower IG to higher IG frequencies (not shown). This energy gain at higher IG frequencies is subsequently balanced by $S^\mathrm{NL}_\mathrm{ig,ig,ss}$, which transports the IG flux to SS frequencies.

\subsection{Infragravity wave dissipation}\label{sec:Results_IGdissipation}
For both slopes, a strong increase followed by an intense reduction of the energy flux occurred at the IG frequencies. To quantify the contribution of the different processes that contributed to the net gain and loss of IG flux inside the surfzone, we cross-shore integrated the IG energy balance terms (only considering cells that were always wet) over the region where the individual terms were either positive or negative.

\begin{figure}
\captionsetup{singlelinecheck = false, justification=justified}
  \centering
    \includegraphics[scale=0.9]{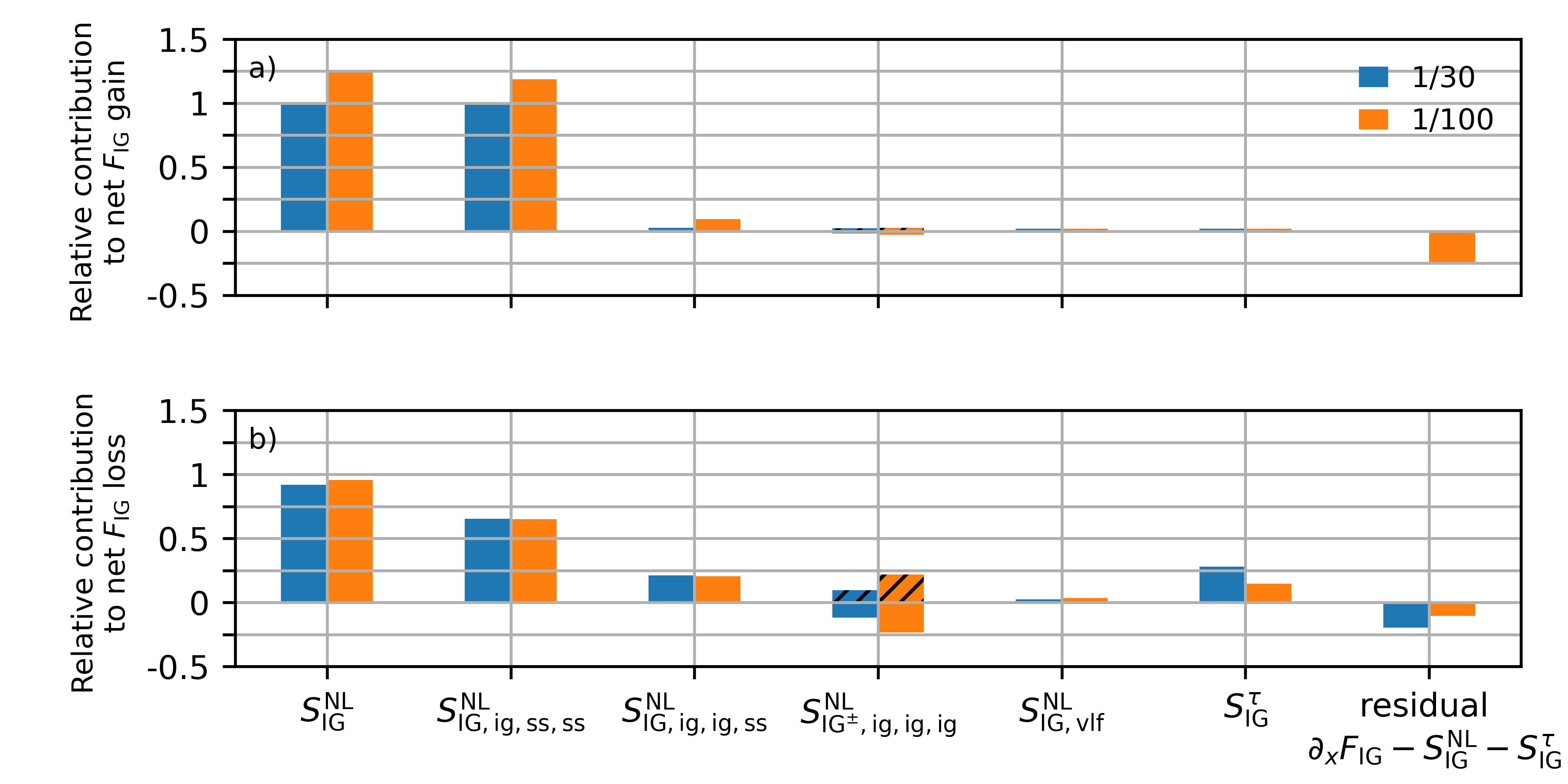}
    \caption{Relative contribution of different energy balance terms to the IG energy flux (a) gain and (b) loss on a 1/30 (blue) and 1/100 (orange) slope.
     Triad interactions among three IG waves (IG-IG-IG) integrate to zero. Instead, Loss terms were cross-shore integrated over cells with  negative values, and gain terms were integrated over positive cells (subscript $\mathrm{IG}^-$ and $\mathrm{IG}^+$, respectively).
    }\label{fig:IGdissipation}
\end{figure}

The net nearshore gain of IG flux was about $2$ times larger for the 1/100 slope than the 1/30 slope ($1.71\;\mathrm{m}^4\mathrm{s}^{-3}$ versus $0.99\;\mathrm{m}^4\mathrm{s}^{-3}$). This gain was nearly completely explained by nonlinear interactions (Fig. \ref{fig:IGdissipation}a), although $S_\mathrm{IG}^\mathrm{NL}$ over estimated the net gain by about $25\%$ on the mild slope (i.e., a net residual of $\approx-25\%$). For both slopes, the nonlinear interactions were dominated by interactions between one IG and two SS components, associated with the well known forcing of IG waves by SS wave groups.

The net nearshore loss of IG flux was about $6$ times larger for the 1/100 slope than the 1/30 slope (-1.93 versus -0.32 $\mathrm{m}^4\mathrm{s}^{-3}$). For the mild slope, the net loss exceeded the net gain (-1.93 versus 1.71 $\mathrm{m}^4\mathrm{s}^{-3}$). This mismatch can be explained by the incoming flux contribution from IG waves generated at the wavemaker (estimated as $F^\mathrm{L^+}_\mathrm{IG}+F^\mathrm{NL}_\mathrm{IG}=0.28 \mathrm{m}^4\mathrm{s}^{-3}$), indicating near complete dissipation of shoreward propagating IG motions on the mild slope that is consistent with negligible reflections at IG frequencies (Fig. \ref{fig:Example_IG}b). For the 1/30 slope, the net loss was only a fraction of the net gain (-0.32 versus 0.99 $\mathrm{m}^4\mathrm{s}^{-3}$), associated with non-negligible reflection of IG waves.  However, the cross-shore integration was limited to cells that were always wet and additional dissipation that occurred in shallower water (including the swash zone) is missing from this analysis and will contribute to this mismatch.

On both slopes, the nearshore loss of the IG flux was nearly completely explained by nonlinear interactions $S_\mathrm{IG}^\mathrm{NL}$ and bottom friction $S_\mathrm{IG}^\tau$ (Fig. \ref{fig:IGdissipation}b), although the combined effect of $S^\mathrm{NL}$ and $S_\mathrm{IG}^\tau$ overestimated the net IG flux loss resulting in a residual of $\approx10-20\%$ (Fig. \ref{fig:IGdissipation}b). For both slopes, bottom friction accounted for at most $\approx25\%$ of the nearshore IG flux losses and nonlinear interactions were responsible for the majority of the flux loss. Separating the nonlinear interactions into correlations between different combinations of VLF, IG and SS wave components (Section \ref{sec:Results_Contr}) indicates that interactions between one IG and two SS components ($S^\mathrm{NL}_\mathrm{IG,ig,ss,ss}$) resulted in the largest transfer of energy away from IG frequencies. The relative contribution of these interactions was comparable for both slopes, and they explained up to $\approx 65\%$ of the nearshore loss in IG flux. Interactions between one IG and two SS components ($S^\mathrm{NL}_\mathrm{IG,ig,ig,ss}$) were typically weaker but still significant on both slopes ($\approx 20 \%$). Energy transfers to frequencies below the IG band ($S^\mathrm{NL}_\mathrm{IG,vlf}$) were small on both slopes.  Consistent with interpretation of laboratory experiments \citep[e.g.,][]{Baldock2000,Baldock2006,Baldock2012} and results from weakly nonlinear energy balances \citep[e.g.,][]{Thomson2006,Bakker2015,DeBakker2016}, these results confirm that interactions with two IG components became significant inside the inner surfzone and result in substantial energy transfer from IG to SS frequencies, where the wave energy is ultimately dissipated.


\section{Discussion}\label{sec:Discussion}
\subsection{Comparison NLSWE energy balance with existing theories}

Theories for nonlinear interactions range from Boussinesq models for resonant waves in shallow water \citep[e.g.,][]{Herbers1997} up to more generalized theories that account for full linear dispersive effects \citep[e.g.,][]{Bredmose2005WaveTransfer,Janssen2006NonlinearTopography}. Several previous studies used the nonlinear interaction term based on the Boussinesq scaling from \citet{Herbers1997}, denoted as $S^\mathrm{NL}_\mathrm{HB97}$, to evaluate the nearshore IG energy balance \citep[e.g.,][]{Thomson2006,Bakker2015,DeBakker2016}. Assuming weakly nonlinear waves by retaining terms up to $\mathcal{O}\left(\delta^3\right)$, and relating $u$ to $\eta$ using $u=\frac{c}{d}\eta$ and $\frac{\partial \eta}{\partial x}=-\frac{1}{d}\frac{\partial \eta}{\partial t}$ with $c=\sqrt{g d}$ (which combine to the linearized flat-bed assumption, $\frac{\partial u}{\partial x}=-\frac{1}{c}\frac{\partial \eta}{\partial t}$), the nonlinear interaction $S^\mathrm{NL}$ term from the NLSWE balance reduces to $S^\mathrm{NL}_\mathrm{HB97}$ (Appendix \ref{sec:A-B}).

The energy balance with Boussinesq scaling of  \citet{Herbers1997} is
\begin{equation}
\frac{\partial F_{f,\mathrm{HB97}}}{\partial x}+S^\mathrm{NL}_{f,\mathrm{HB97}}=S^\tau_f,
\end{equation}
in which $F_{f,\mathrm{HB97}}=E c_g$ and $S^\mathrm{NL}_{f,\mathrm{HB97}}$ depends on a coupling coefficient and Q, the imaginary part of the bispectrum. We computed $S^\mathrm{NL}_\mathrm{HB97}$ following \citet{Henderson2006},
\begin{eqnarray}
  S^\mathrm{NL}_{f,\mathrm{HB97}}=\frac{3 g \omega}{2 h} Q_f \left( \eta^2, \eta \right)\!,
\end{eqnarray}
where $h$ is the mean water depth ($h=d+\overline{\eta}$). The flux term $F$ of the Boussinesq scaled energy balance breaks down in the case of significant shoreline reflections (as is the case on the 1-30 slope, Fig. \ref{fig:Example_IG}a). As an alternative, we used the linear flux estimates to evaluate this term, which yields similar results in the case of weak shoreline reflections (Fig. \ref{fig:Example_IG}b).

\begin{figure}
\captionsetup{singlelinecheck = false, justification=justified}
    \centering
    \includegraphics[scale=0.9]{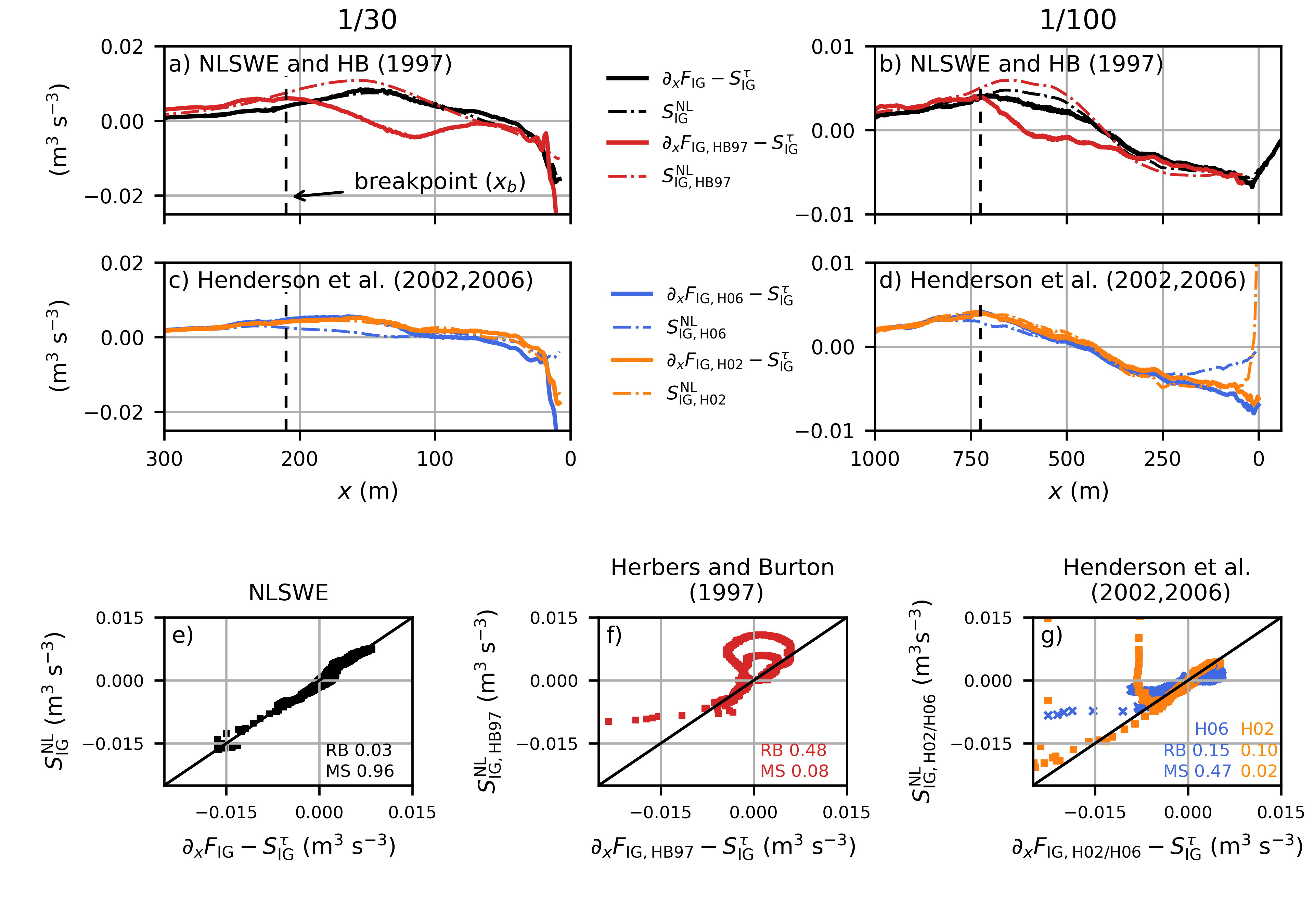}
    \caption{Nearshore IG energy balance for different theories on (left) 1/30 and (right) 1/100  slope.  Top panels (a-d) show the cross-shore evolution of the cross-shore energy flux gradient minus the bottom stress ($\partial_x F_\mathrm{IG}-S^{\tau}_{\mathrm{IG}}$) and  the nonlinear interaction term
   ( $S^{\textrm{NL}_\mathrm{IG}}$) of (a-b) NLSWE balance (this paper) and Boussinesq scaled balance \citet{Herbers1997}, and (c-d) balance of \citet{Henderson2002} and \citet{Henderson2006}.
     Bottom panels (e-g) are scatter plots of  $S^\textrm{NL}_\mathrm{IG}$ versus $\partial_x F_\mathrm{IG}-S^{\tau}_{\mathrm{IG}}$ for
    (e) the NLSWE balance (f)  \citet{Herbers1997}, and (g) \citet{Henderson2002} and \citet{Henderson2006}. In each panel, the relative bias and skill scores are given for the respective theory. Solid lines indicate closure of the IG energy balance. For the Boussinesq scaled balance, results are not plotted when the underlying WKB assumption is violated ($\frac{\omega^2 d}{g}<10 h_x^2$, with $\omega$ from $T_{m01,\mathrm{IG}}$).}
    \label{fig:Comparison}
\end{figure}

The Boussinesq scaled energy flux and nonlinear interaction term are in balance in the shoaling region ($x>x_b$, Fig. \ref{fig:Comparison}a-b), which confirms that in this region the weakly nonlinear and dispersive balance closes to the order considered. However, this balance deteriorated significantly in the surfzone on both slopes, where wave nonlinearity (amplitude over depth) is large. Here,
$S^\mathrm{NL}_\mathrm{IG,HB97}$ over estimated the linear flux gradient (Fig. \ref{fig:Comparison}a-b). In contrast, the fully nonlinear but non-dispersive energy balance based on the NSLWE closed with much smaller residual compared to the Boussinesq scaled energy balance throughout the surfzone (compare Fig. \ref{fig:Comparison}e and \ref{fig:Comparison}f).

\citet{Henderson2002} derived a nonlinear energy balance for IG wave frequencies that considers the mean potential and total pseudo-kinetic energy derived from the Lagrangian mean energy in the water column (see Appendix \ref{sec:A-C}). \citet{Henderson2006} subsequently adapted this theory by assuming weak nonlinearity. The weakly nonlinear theory of \citet{Henderson2006} did not provide a closing balance between the energy flux gradient $\partial_x F_\mathrm{IG,H06}$ and nonlinear interaction term $S^\mathrm{NL}_\mathrm{IG,H06}$ (Fig. \ref{fig:Comparison}c-d and \ref{fig:Comparison}g). The residual was largest in the inner surfzone, where the assumption of weak nonlinearity breaks down (e.g., $S^{\mathrm{NL}}_\mathrm{ig,ig,ss}$ becomes non-negligible, Fig. \ref{fig:Example_IG}g-h), and $S^\mathrm{NL}_\mathrm{H06}$ under estimated the loss of IG energy flux. The nonlinear version of this balance \citep{Henderson2002} describes a closer balance between the flux gradient $\partial_x F_\mathrm{H02}$ and nonlinear interaction term  $S^{\mathrm{NL}}_\mathrm{H02}$ throughout most of the surfzone (Fig. \ref{fig:Comparison}c-d). However,  $S^\mathrm{NL}_\mathrm{H02}$ blows up close to the shoreline (where $d\rightarrow0$ m) due to its dependence on the Lagrangian velocity ($\tilde{u}=u+\frac{1}{d}\eta u$, Appendix \ref{sec:A-C}).

\subsection{Infragravity wave breaking}

Past studies on mild slopes have linked substantial IG energy dissipation with IG breaking \citep[e.g.,][]{VanDongeren2007,Bakker2014,Bakker2015,DeBakker2016}. Combining energy balances with bispectral analysis, several previous studies found that significant transfers of IG energy to higher harmonics coincided with skewed and asymmetric shapes at the IG frequencies (as was observed in the time signals on the 1/100 slope, Fig. \ref{fig:Timestacks}) and linked this to the possibility that IG waves were breaking \citep[e.g.,][]{Bakker2015,DeBakker2016}. These studies relied on weakly nonlinear energy balances that do not close at IG frequencies, complicating a quantitative understanding of the nearshore IG wave dissipation. The fully nonlinear non-dispersive energy balance derived in this work did (nearly) completely close at IG wave frequencies, and allowed for a meticulous study into the nearshore dissipation of IG waves (Section \ref{sec:Results_IGdissipation}).

\begin{figure}
\captionsetup{singlelinecheck = false, justification=justified}
    \centering
     \includegraphics[scale=0.9]{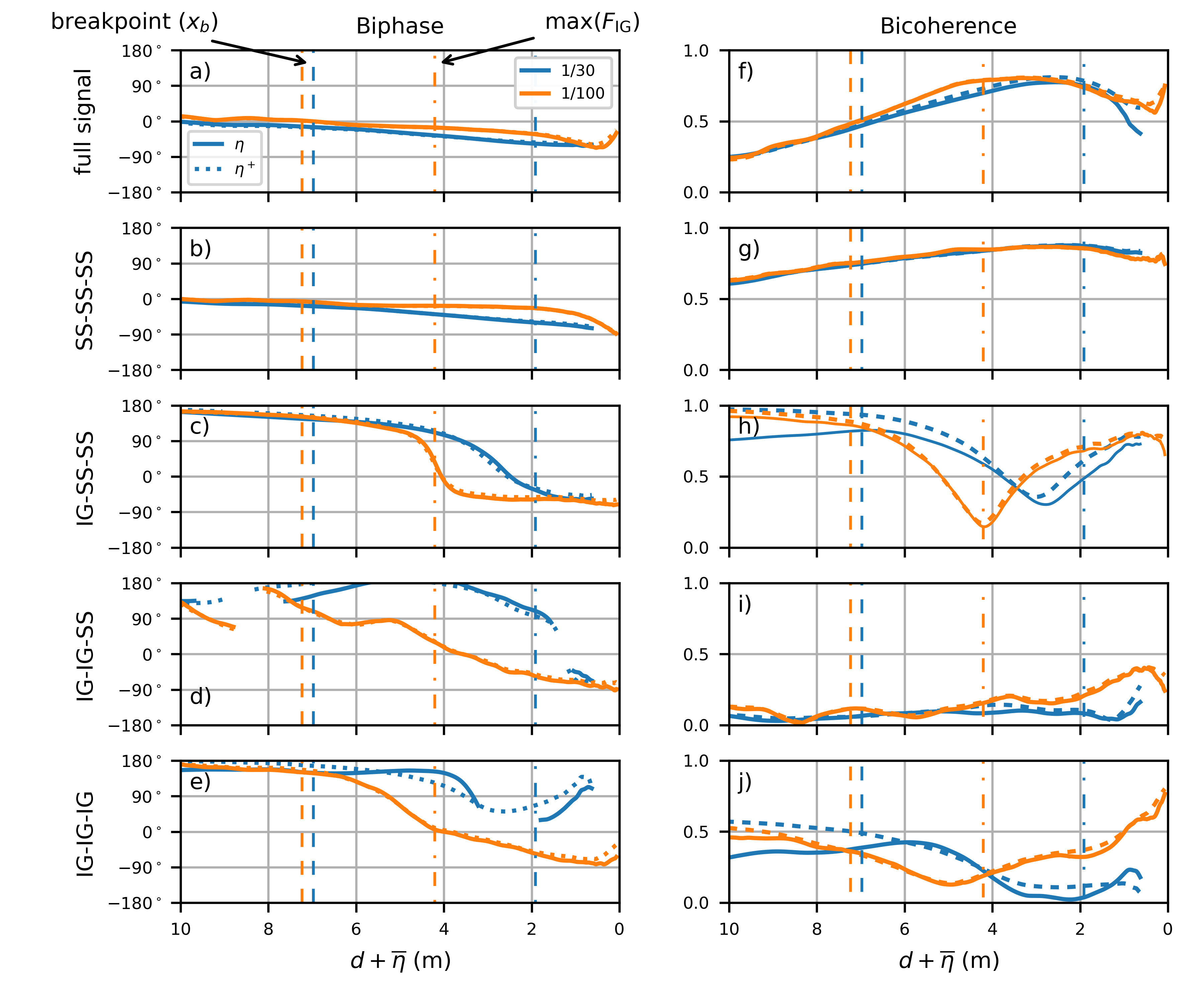}
   \caption{Cross-shore variation of the (left) biphase $\beta$ and (right) bicoherence $C$  versus total depth for (a,b) the full signal (all frequencies), and (b-j) triads between different combinations of IG and SS components based on the total signal (full line) and signal corresponding to shoreward propagating waves (dotted line) for the 1/30 (blue lines) and 1/100 (orange lines) slope. Vertical lines indicate the location of (dashed) the seaward surfzone edge and (dash-dotted) where the IG flux gradient $\partial_x F$ changes sign (see Fig. \ref{fig:Example_IG}). To suppress noise, biphases are not shown when the corresponding bicoherence $<0.05$.
 }
  \label{fig:biphase}
\end{figure}

The IG wave dynamics on the 1/30 and 1/100 slope showed remarkably similar patterns up to the inner surfzone. In the shoaling region, energy was continuously transferred from SS wave groups to the IG band by nonlinear interactions (Fig. \ref{fig:Example_IG}e-f). As the waves approach the surfzone, the biphase between these IG-SS-SS triads gradually moved away from the equilibrium response $\beta_\mathrm{ig,ss,ss}=180^\circ$ (Fig. \ref{fig:biphase}c). The IG band received energy through the first part of the outer surfzone (Fig. \ref{fig:Example_IG}e-f), up to the location where $\beta_\mathrm{ig,ss,ss}\approx0^\circ$ (Fig. \ref{fig:biphase}c), indicating that here the largest SS waves propagate on the crest of the IG waves (see the time signals near the dashed red line in Fig. \ref{fig:Timestacks}). Further shoreward, the IG band lost energy to the SS band due to $S^\mathrm{NL}_\mathrm{ig,ss,ss}$ interactions. These interactions persisted until deep inside the surfzone, and were responsible for $\approx65\%$ of the net IG flux loss (Fig. \ref{fig:IGdissipation}b). In the inner surfzone, nonlinear interactions between two IG and one SS harmonic became significant on both slopes (Fig. \ref{fig:Example_IG}g-h), accounting for approximately $20\%$ of the net IG flux loss (Fig. \ref{fig:IGdissipation}b).

Only on the 1/100 slope, triad interactions involving at least two IG components became increasingly significant in the inner surfzone ($x<250$ m), with increasing bicoherence (Fig. \ref{fig:biphase}i,j). The biphase of these IG triads and of the other triads involving at least one IG harmonic tended to $\beta=-90^\circ$, indicative of a forward-pitching (saw-tooth shaped) wave  ($0 < d < 1$ m, Fig. \ref{fig:Timestacks}). Surprisingly, the IG-IG-SS triads only accounted for approximately $25\%$ of the net IG flux loss, as $S^\mathrm{NL}_\mathrm{ig,ss,ss}$ drained the majority of the IG flux (Fig. \ref{fig:IGdissipation}b). Seaward of the inner surfzone, $S^\mathrm{NL}_\mathrm{ig,ss,ss}$ resembled interactions between IG waves and SS groups that caused an energy transfer from the IG waves back to the SS waves. In the inner surfzone IG waves have a saw-tooth shape (Fig. \ref{fig:Timestacks}) and such IG-SS-SS triads can also represent IG self interactions. Although we cannot establish which of these two alternatives are responsible for the energy transfer, we expect that IG self interactions may dominate $S^\mathrm{NL}_\mathrm{ig,ss,ss}$ when the water depth is sufficiently small ($d<1$ m, Fig. \ref{fig:bulk}d). Seaward of this location, $S^\mathrm{NL}_\mathrm{ig,ss,ss}$ accounted for $\approx46\%$ of the total IG flux loss (equivalent to $\approx70\%$ of the loss induced by IG-SS-SS triads). This indicates that even though IG waves were breaking on the mild slope, nonlinear interactions between IG waves and SS wave groups were responsible for at least half of the net IG flux loss.

\subsection{Validity of NLSWE balance at sea-swell frequencies}
The energy balance based on the NLSWE provided a good description of the surfzone IG dynamics simulated
with the fully nonlinear and dispersive SWASH model.  Despite neglecting dispersive effects in NLSWE (not in SWASH), the IG energy balance approximately closes across the surfzone. IG energy losses do occur, but are mostly explained through nonlinear transfers to the SS band. Residuals at SS frequencies are harder to directly interpret. Outside the surfzone, the small residual (Fig. \ref{fig:Example_SS}c-d) is likely due to the exclusion of work performed by the non-hydrostatic pressure in the flux terms. Inside the surfzone, a combination of neglected physics in the analysis (i.e., turbulent stresses) and dissipative losses due to breaking are responsible for the residual.

In SWASH, breaking is modeled as a discontinuity and dissipation occurs at the scale of the spatial resolution $\Delta x$. The associated frequency scale of O($\sqrt{gd}/\Delta x$) falls within the superharmonic range ($f>3f_{p}$) throughout the domain. As a consequence, the residual in the primary SS band is likely primarily the result of inaccuracies in the NLSWE derived balance (Fig. \ref{fig:Example_SS}c-d). Although we suspect that the large residual in the superharmonic range is a dissipation signature, the required extension to three-dimensional non-hydrostatic flow is outside the current scope of work. We further found that inaccuracies in the NLSWE balance become large when spatial and temporal resolution are mismatched in the analysis. For example, when a high spatial resolution captures the dissipation at superharmonics but superharmonics are not resolved due to a low temporal sampling resolution (Appendix \ref{sec:A-D}).

\section{Conclusions}\label{sec:Conclusions}
A fully nonlinear non-dispersive energy balance for surfzone waves based on the nonlinear shallow water equations was derived and applied to unidirectional spectral waves on a moderately and a mildly sloping beach. The energy balance was evaluated based on output from a fully nonlinear and dispersive wave model (the non-hydrostatic wave-flow model SWASH).

The new theory predicts a closed energy balance at the IG frequencies. Throughout the shoaling region and most of the surfzone, the gain and loss of the IG energy flux was primarily balanced by nonlinear wave interactions. The nonlinear interactions explained the energy growth through the shoaling region up to the outer surfzone, and also caused most of the energy loss deeper inside the surfzone. Bottom friction caused only small dissipation in very shallow water.
The nonlinear energy transfer was primarily explained by interactions between a single IG and two SS wave components. Only deep inside the surfzone on the mildly sloping beach, interactions between two IG and a single SS wave component (suggesting IG wave breaking) became significant and contributed to the IG energy flux loss.

Existing theories to estimate the nearshore IG energy balance that assume weak nonlinearity do not account for all nonlinear contributions to the energy flux and/or nonlinear interaction term. This degrades the energy balance closure and complicates the interpretation of the IG energy balance. The fully nonlinear non-dispersive balance was shown to provide an improved description of the IG energy balance inside the surfzone.

\backsection[Acknowledgements]{R. Guza was supported by USACE and California State Parks. P. Smit was supported by the Office of Naval Research through grants N00014-20-1-2439. We are very grateful to Julia Fiedler for her contributions during the initial phase of this work.}

\backsection[Declaration of interests]{The authors report no conflict of interest.}

\appendix

\section{Sensitivity of energy balance to sampling in space and time}\label{sec:A-D}

\begin{figure}
\captionsetup{singlelinecheck = false, justification=justified}
    \centering
    \includegraphics[scale=0.9]{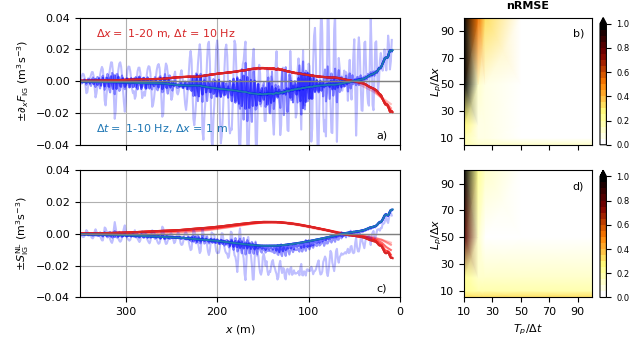}
    \caption{Sensitivity of the bulk IG energy balance terms to sampling resolution in space ($\Delta x$) and time ($\Delta t$) on 1/30 slope. Cross-shore variation of (a) energy flux gradient and (b) nonlinear interaction for a varying grid size given a fixed time sampling ($\Delta t=10$ Hz, $T_p/\Delta t$=100, red lines), and a varying time sampling given a fixed grid size ($\Delta x=1$ m, $ L_p/ \Delta x$= 100, blue lines). Light to dark colors indicate coarse to fine sampling resolutions, and the result for a varying time step are mirrored around the y-axis for visualization. The right panels show the normalised root-mean-square-error (nRMSE) for the full parameter space, with the spatial and temporal sampling related to the peak off-shore wave length $L_p$($\approx 100$ m) and wave period $T_p$($\approx10$ s), respectively.}
    \label{fig:EB_sensitivity}
\end{figure}

The spatial and temporal sampling required to estimate accurately the nearshore IG energy balance was explored by evaluating the energy balance terms with variable spatial ($\Delta x=1-8$ m) and temporal  ($\Delta t = 2-10$ Hz ) resolution
(The SWASH numerical model output is $\Delta x=1$ m and $\Delta t = 40$ Hz). For a constant sampling resolution of $\Delta t=10$ Hz, the bulk IG energy flux gradient $\partial_x F_\mathrm{IG}$ and nonlinear interaction term $S^\mathrm{NL}_\mathrm{IG}$ were insensitive to the spatial resolution (red lines in panels a,c of Fig. \ref{fig:EB_sensitivity}). In contrast, estimates of both terms were sensitive to the temporal resolution for a fixed grid resolution of $\Delta x=1$ m (blue lines).  Sampling errors were quantified with the normalised root-mean-square error (nRMSE)
\begin{equation}
    \mathrm{nRMSE}=\frac{ \sqrt{ \left\langle \left( Q - Q_r \right)^2  \right\rangle }}{ \left|Q_r\right|_\mathrm{max} },
\end{equation}
where angular brackets indicates averaging over all cross-shore output locations, $\left|...\right|_\mathrm{max}$ indicates the maximum absolute value, and $Q$ is a term of the bulk IG energy balance (with subscript r indicating the reference results corresponding to the finest spatial and temporal sampling). Errors are generally small when space and time resolution are relatively high
$L_p/\Delta x, T_p/\Delta t > 30$ (Fig. \ref{fig:EB_sensitivity}). Errors in IG flux gradients and noninear terms are largest when a high spatial resolution  ($L_p/\Delta x >20, \Delta x<5$ m) is mismatched with low temporal resolution ($T_p/\Delta t<20, \Delta t>0.5 s$).

\section{Contributors to the nonlinear energy flux and nonlinear interaction term}\label{sec:A-A}

Bispectral analysis can be used to estimate the contributions of correlations between particular wave components to third order wave statistics An alternative to the bispectrum \citep{Bakker2015} is time-domain analysis (band-passed correlations) between filtered signals
\citet{Fiedler2019TheModeling}. Using time domain band-passing,  a wave or velocity signal $X(t)$ is decomposed into VLF, IG, and SS motions

\begin{equation}
    X=X_\mathrm{vlf}+X_\mathrm{ig}+X_\mathrm{ss},\label{eq:B1}
\end{equation}

Substitution of Eq. \eqref{eq:B1} in the general co-spectrum between three signal $X$, $Y$ and $Z$ yields,
\begin{equation}\label{eq:C_XYZ}
\begin{aligned}
\mathrm{C}_f\! \left( X Y; Z \right) &= \mathrm{C}_f\! \left( \left(X_\mathrm{vlf}+X_\mathrm{ig}+X_\mathrm{ss}\right) \left(Y_\mathrm{vlf}+Y_\mathrm{ig}+Y_\mathrm{ss}\right); \left( Z_\mathrm{vlf}+Z_\mathrm{ig}+Z_\mathrm{ss} \right)\right) \\
&= \mathrm{C}_f\! \left(X_\mathrm{ss} \left(Y_\mathrm{vlf}+Y_\mathrm{ig}+Y_\mathrm{ss}\right); \left( Z_\mathrm{vlf}+Z_\mathrm{ig}+Z_\mathrm{ss} \right)\right) \\
&+\mathrm{C}_f\! \left( X_\mathrm{ig} \left(Y_\mathrm{vlf}+Y_\mathrm{ig}+Y_\mathrm{ss}\right); \left( Z_\mathrm{vlf}+Z_\mathrm{ig}+Z_\mathrm{ss} \right)\right) \\
&+\mathrm{C}_f\! \left( X_\mathrm{vlf} \left(Y_\mathrm{vlf}+Y_\mathrm{ig}+Y_\mathrm{ss}\right); \left( Z_\mathrm{vlf}+Z_\mathrm{ig}+Z_\mathrm{ss} \right)\right)\!.
\end{aligned}
\end{equation}
Decomposing the nonlinear interaction term $S^\mathrm{NL}_f$ (Eq. \ref{eq:EB-Snl}), and collecting terms that are permutations of the same signals yields
\begin{eqnarray}
  S^{\mathrm{NL}}_f &=&  S^{\mathrm{NL}}_{f,\mathrm{ss,ss,ss}} +S^{\mathrm{NL}}_{f,\mathrm{ig,ig,ig}} + S^{\mathrm{NL}}_{f,\mathrm{ig,ig,ss}} + S^{\mathrm{NL}}_{f,\mathrm{ig,ss,ss}} + S^{\mathrm{NL}}_{f,\mathrm{vlf}},
\end{eqnarray}
in which $S^{\mathrm{NL}}_{f,\mathrm{ig,ig,ig}}$ and $S^{\mathrm{NL}}_{f,\mathrm{ss,ss,ss}}$ represent triad interactions between three IG and three SS wave components, respectively, $S^{\mathrm{NL}}_{f,\mathrm{ig,ig,ss}}$ represents interactions between two IG and one SS wave component, and $S^{\mathrm{NL}}_{f,\mathrm{ig,ss,ss}}$ represents interactions between a single IG and two SS wave components. Finally, $S^{\mathrm{NL}}_{f,\mathrm{vlf}}$ is short for all interactions that include at least one VLF component (the last term in Eq. \ref{eq:C_XYZ}). The nonlinear flux $F^\mathrm{NL}_{f}$ can be decomposed in a similar fashion,
\begin{eqnarray}
  F^{\mathrm{NL}}_f &=& F^{\mathrm{NL}}_{f,\mathrm{ig,ig,ig}} + F^{\mathrm{NL}}_{f,\mathrm{ig,ig,ss}} + F^{\mathrm{NL}}_{f,\mathrm{ig,ss,ss}} + F^{\mathrm{NL}}_{f,\mathrm{ss,ss,ss}} + F^{\mathrm{NL}}_{f,\mathrm{vlf}}.
\end{eqnarray}

\section{Comparison with weakly nonlinear theory}\label{sec:A-B}
We consider unidirectional waves over constant depth with a typical ratio $\delta$ between amplitude and depth that is small but finite. Neglecting friction and retaining  terms up to $O(\delta^3)$ in nonlinearity, the NLSWE energy balance (Eq. \ref{eq:enbalance}) can be written as,
\begin{equation}
    \frac{\partial E_f}{\partial t} + \frac{\partial F_f}{\partial x} = S^\mathrm{nl}_f\label{eq:C1},
\end{equation}
with
\begin{eqnarray}
  E_f &=&   \frac{g}{2}\mathrm{C}_f\!(\eta,\eta) + \frac{1}{2} \mathrm{C}_f\!(Du,u), \label{eq:C2} \\
  F_{f} &=& g \mathrm{C}_f\! \left( D u ; \eta \frac{}{} \right)
  + \frac{1}{2} \mathrm{C}_f\! \left( d u^2; u \frac{}{} \right), \label{eq:C3}\\
  S_f^{\textrm{NL}} &=& \frac{1}{2} \left[ \mathrm{C}_f\!\left( d u^2; \frac{\partial u}{\partial x} \right) - \mathrm{C}_f\!\left( d u; u \frac{\partial u}{\partial x} \right) \right] \label{eq:C4}\\
  &+& \frac{1}{2} g \left[ \mathrm{C}_f\! \left( \eta u; \frac{\partial\eta}{\partial x_m} \right) - \mathrm{C}_f\! \left( u; \eta \frac{\partial\eta}{\partial x} \right) \right]\!.\nonumber
\end{eqnarray}
With waves propagating in the positive x direction, the negative Riemann characteristic of the 1D nonlinear shallow water equations has the constant value of $u-2\sqrt{gD}=-2c$ (with $c=\sqrt{gd}$). The nonlinear depth averaged velocity and surface elevation are related by
\begin{equation}
    u = 2\sqrt{gD} - 2c \approx \frac{c}{d}\left[\eta - \frac{\eta^2}{2d} + O(\delta^2) \right]\!.
    \label{eq:C5}
\end{equation}
Using Eq. \eqref{eq:C5} and retaining terms through $O(\delta^3)$ the energy and flux (Eq. \ref{eq:C2} and \ref{eq:C3}, respectively) reduce to
\begin{equation}
    E_f =  g \mathrm{C}_f\! \left( \eta ; \eta \frac{}{} \right)\!,
\end{equation}
\begin{equation}
  F_{f}= gc\left[ \mathrm{C}_f\! \left( \eta ; \eta\frac{}{} \right)+   \frac{1}{d}\mathrm{C}_f\! \left( \eta ; \eta^2 \frac{}{}    \right) \right]= c\left[ E_f + \frac{g}{d} \int \mathcal{B}^{\mathrm{Re}}_{f-f^\prime,f^\prime} \mathrm{d}f^\prime \right]\!.
  \label{eq:weaknlflux}
\end{equation}
The real part of the surface elevation bispectrum $\mathcal{B}^{\mathrm{Re}}_{f-f^\prime,f^\prime}$ is typically neglected in weakly nonlinear balances, under the assumption mean quantities change on a slow spatial scale. Consequently gradients ($\frac{\partial F_f}{\partial x}$) are of order O($\delta$), so the flux correction enters dynamically at O($\delta^4$) in Eq. \eqref{eq:C1}, and can be neglected  \citep{Herbers1997}. Lastly, to the leading order in  nonlinearity and neglecting bottom slope $\partial_x \approx -\frac{1}{c}\partial_t$ and using that in the frequency domain $\partial_t \rightarrow \mathrm{i}2\pi f$, yields
\begin{equation}
  S_f^{\textrm{NL}} = \frac{gc}{d} \left[ \mathrm{C}_f\!\left( \eta^2; \frac{\partial \eta}{\partial x} \right) - \mathrm{C}_f\!\left( \eta; \eta \frac{\partial \eta}{\partial x} \right) \right] = -\frac{ 3\omega gc}{2d}
  \int \mathcal{B}^{\mathrm{Im}}_{f-f^\prime,f^\prime} \mathrm{d}f^\prime\!,
  \label{eq:weaknlsource}
\end{equation}
with $\mathcal{B}^{\mathrm{Im}}_{f-f^\prime,f^\prime}$ the imaginary bispectrum, consistent with previous weakly-nonlinear WKB balances for shoreward propagating waves \citep[e.g.,][]{Herbers1997}.

\section{Comparison to \citet{Henderson2002}}\label{sec:A-C}

The balance considered in \citet{Henderson2002} can be found from combination of the frequency domain conservative momentum and mass equations alone as,
\begin{equation}
     \lim_{\Delta f \rightarrow 0} \frac{1}{\Delta f}\left\langle\left( \overline{\tilde{u}}_m^f/2 \times \mathrm{Eq.}\; \eqref{eq:II}^* + \mathrm{C.C.} \right) + \left( g\overline{\eta}^f/2 \times \mathrm{Eq.}\; \eqref{eq:I}^* + \mathrm{C.C.} \right)\right\rangle\!,
\end{equation}
where $\tilde{u}_m = \frac{d+\eta}{d}u_m$ is the Lagrangian (or mass transport) velocity. The result can be written as a balance of the form,
\begin{equation}
  \frac{\partial \tilde{E}_f}{\partial t} + \frac{\partial }{\partial x_n}\left( \tilde{F}^\mathrm{L}_{n,f}+\tilde{F}^{\mathrm{NL}}_{n,f}\right) = \tilde{S}^{\mathrm{NL}}_f + \tilde{S}^{\tau}_f,
  \label{eq:enbalanceH02}
\end{equation}
in which
\begin{eqnarray}
  \tilde{E}_f &=& \frac{1}{2}g \mathrm{C}_f\! \left( \eta; \eta \frac{}{} \right) + \frac{d}{2} \mathrm{C}_f\! \left(\tilde{u}_m; \tilde{u}_m \frac{}{} \right)\label{eq:D3}\!,\\
  \tilde{F}^{\mathrm{L}}_{n,f} &=& g \mathrm{C}_f\! \left( du_n ; \eta \frac{}{} \right)\!,\\
  \tilde{F}^{\mathrm{NL}}_{n,f} &=& g \mathrm{C}_f\! \left( \eta u_n ; \eta \frac{}{} \right) + \mathrm{C}_f\! \left( D u_n u_m + \frac{\delta_{m,n}}{2}\eta^2; \tilde{u}_n \frac{}{} \right)\!,\\
  \tilde{S}_f^{\textrm{NL}} &=& \mathrm{C}_f\!\left(  D u_n u_m + \frac{\delta_{m,n}}{2}\eta^2; \frac{\partial \tilde{u}_n}{\partial x_m} \frac{}{}  \right)\!,\\
  \tilde{S}_f^{\tau} &=& -\mathrm{C}_f\! \left( \tilde{u}_m; \tau_m \right)\!.\label{eq:D7}
\end{eqnarray}
Though qualitatively similar to the NLSWE energy balance (Eq. \ref{eq:enbalance}), this is not an energy balance. Integrating Eqs. \eqref{eq:D3}-\eqref{eq:D7} over all frequencies, and manipulating the result to resemble closely the bulk energy balance (Eqs. \ref{eq:encons}-\ref{eq:bulktaub}), results in
\begin{equation}
\frac{\partial}{\partial t}\left[ E +  \left\langle\frac{\eta}{d}D u_m u_m\right\rangle\right] + \frac{\partial F_n}{\partial x_n}  = S^\tau - \left\langle \frac{\eta u_m \tau_m}{d} - \frac{\eta u_m}{d}\frac{\partial D u_m u_n}{\partial x_n} - g \frac{\tilde{u}_m}{2} \frac{\partial \eta^2}{\partial x_m}\right\rangle\!.
\end{equation}
In effect, this balance considers the time evolution of a quantity that is the sum of mean potential and a total pseudo-kinetic energy derived from the Lagrangian mean energy in the water column. This differs from the standard energy measure by the term in brackets on the left hand side. The above form (spectral or bulk) is non-unique, and the right hand side terms can be manipulated into flux-like contributions and source terms through the chain rule. This is reflected in the different forms of the balance considered in, for example, \citet{Henderson2002} and \citet{Henderson2006}. However, in all cases the bulk balance takes a non-conserved form, and is generally not strictly interpretable as an energy balance. That said, for mild slopes and mild nonlinearity the balance is consistent with weakly nonlinear theory \citep{Henderson2006}, and significant differences are only expected at high Ursell numbers.

\clearpage

\clearpage

\bibliographystyle{jfm}

\end{document}